\documentclass[12pt]{article}

\usepackage{amsmath}
\usepackage{amsfonts,dsfont,mathrsfs}
\usepackage{amssymb}
\usepackage{graphics,graphicx,tikz,tensor}
\usetikzlibrary{arrows,decorations.pathmorphing,patterns}
\usepackage[linktocpage]{hyperref}
\hypersetup{
	colorlinks=true,
	linkcolor=blue,
	filecolor=magenta,      
	urlcolor=blue,
	citecolor=magenta
}
\numberwithin{equation}{section} 
\usepackage{caption,subcaption}
\usepackage{cancel}
\usepackage{tabu}
\usepackage{empheq}

\usepackage[noadjust]{cite}

\def\be{\begin{equation}}
	\def\ee{\end{equation}}
\def\bea{\begin{align}}
	\def\eea{\end{align}}
\def\beaq{\begin{eqnarray}}
	\def\eeaq{\end{eqnarray}}

\usepackage{color}

\begin{document}

\hfill
	\vspace{18pt}
	\begin{center}
		{\Large 
			\textbf{
                          Fortuity beyond counting: an explicit construction
                        }}
		
	\end{center}

		\vspace{8pt}
	\begin{center}
		{\textsl{Stefano Giusto}}
		\vspace{.2cm}

		{\small Dipartimento di Fisica,  Universit\`a di Genova, Via Dodecaneso 33, 16146, Genoa, Italy} \\  \vspace{0pt}
		
		{\small I.N.F.N. Sezione di Genova,
			Via Dodecaneso 33, 16146, Genoa, Italy}\\
			
			\vspace{.5cm}

	{\textsl{James Inglis$^{\;a}$\, and\, Rodolfo Russo$^{\;b}$}}
		\vspace{.2cm}
				
		{\small ${}^{a}$ Department of Physics and Astronomy and ${}^{b}$ School of Mathematical Sciences, \\ Queen Mary University of London, Mile End Road, London, E1 4NS, United Kingdom} \\ 
					
	\end{center}

	\vspace{12pt}

	\begin{center}
		\textbf{Abstract}
	\end{center}
	
	\vspace{4pt} 
	\begin{center}
	\begin{minipage}{15.2cm}
	{\small
	\baselineskip=15pt
	\parskip=5pt
	\noindent 
We reconsider the ``fortuity'' mechanism in the D1D5 CFT focusing on the K3 symmetric orbifold. Going beyond the counting of BPS states, we investigate perturbatively how the explicit form of the BPS cohomologies is modified by the twist-two deformations. We calculate the action of the supercharges in the sector $(h,j)=(1,0)$ for different values of the central charge and derive explicit expressions for the primary states. Equipped with this information, we compare some protected three-point couplings in the free and the gravity regime. We show that agreement between the two descriptions imposes non-trivial constraints on the identification of monotone and fortuitous states. In particular, we argue that the map relating theories with different values of the central charge must and can be defined so as to commute with the supercharges that define the cochain complex. We then study the three-point correlators between the fortuitous and monotone states identified in our analysis to assess whether the two sectors are dynamically decoupled. We find an explicit example of a non-vanishing coupling between two monotone and a fortuitous state, providing evidence that the two sectors are dynamically connected.
		}
	\end{minipage}
	\end{center}	
		
		\vspace{1cm}

		\thispagestyle{empty}

		\vfill
		\vskip 5.mm
		\hrule width 5.cm
		\vskip 2.mm
		{
			\noindent {\scriptsize e-mails: j.inglis@qmul.ac.uk, stefano.giusto@ge.infn.it, r.russo@qmul.ac.uk}
		}

		\setcounter{footnote}{0}
		\setcounter{page}{0}

		
		\baselineskip=17pt
		\parskip=5pt
		
		\newpage

{		
\hypersetup{linkcolor=black}
\setcounter{tocdepth}{3}
}

\vspace{.7cm}

\section{Introduction}

In holographic theories of gravity~\cite{Maldacena:1997re} the microstates of a black hole are described very concretely in the dual CFT by local operators that have conformal dimensions of the order of the central charge $c$. The spectrum of such ``heavy'' operators is believed to be very complicated and a signature of this on the bulk side is that black hole physics is chaotic~\cite{Shenker:2013pqa}. It is more challenging to understand the chaotic features of black holes on the CFT side, in particular in the context of an explicit top-down AdS/CFT duality like ${\cal N}=4$ Super Yang-Mills (SYM) or the D1D5 CFT. A proposal is that the spectrum of heavy operators depends very erratically on the central charge of the theory and in general it is not possible to define the $c\gg 1$ limit for {\em individual} heavy operators~\cite{Schlenker:2022dyo}. The chaotic nature of black hole physics is relevant even for supersymmetric examples in string theory that are well studied in the context of the AdS/CFT duality~\cite{Chen:2024oqv}.

Looking at such supersymmetric cases, it has been understood that there exists a class of heavy states that is tractable and well behaved even in the large $c$ limit: these are multi-particle operators made by a large, ${\cal O}(c)$, number of mutually BPS gravity particles, i.e. Chiral Primary Operators (CPOs) and their descendants with respect to the anomaly-free part of the superconformal algebra. Since, under the AdS/CFT dictionary, each constituent corresponds on the bulk side to an elementary excitation in the supergraviton multiplet, these objects are often called graviton gas states. It is possible to form coherent superpositions of such states to find solitonic objects that correspond to smooth, regular solutions on the gravity side and at the same time have a precise CFT dual: prototypical cases are the $1/2$-BPS solutions found in~\cite{Lin:2004nb} for the AdS$_5$/CFT$_4$ duality and in~\cite{Lunin:2001fv,Kanitscheider:2007wq} in the AdS$_3$/CFT$_2$ setting. 

In this paper we focus on the D1D5 CFT which is dual to type IIB string theory compactified on AdS$_3 \times$ S$^3 \times$ K$3$. This setup is particularly well-suited to the study of BPS black holes, since it is sufficient\footnote{In the ${\cal N}=4$ SYM theory the supersymmetric regular black holes of \cite{Gutowski:2004ez,Gutowski:2004yv} only preserve $1/16$ of the supersymmetries.} to consider an ensemble of $1/4$-BPS heavy states in the CFT to have a large degeneracy accounting for the Bekenstein-Hawking entropy of a large black hole~\cite{Strominger:1996sh}.   Heavy graviton gas states in this sector are known explicitly and go under the name of ``superstrata''~\cite{Bena:2015bea,Bena:2017xbt}, see~\cite{Shigemori:2020yuo} for a review. As their $1/2$-BPS analogue, superstrata are regular, horizonless solutions of the supergravity equations.  Even if these states preserve the same supersymmetries of a large black hole and can approximate its geometry arbitrary well~\cite{Bena:2016ypk}, they do not account for the Bekenstein-Hawking entropy (not even the scaling in $c$ is the one expected). Thus the vast majority of heavy states comes from a different class of states which has been recently dubbed ``fortuitous''~\cite{Chang:2024zqi}.

Finding explicit fortuitous states is hard~\cite{Chang:2013fba} even when working at small values of $c$ and close to the free CFT, i.e. far away from the gravity regime; see~\cite{Chang:2022mjp,Choi:2022caq} for the first example in the context of ${\cal N}=4$ SYM. The existence of a {\em mechanism} (trace relations in ${\cal N}=4$) responsible for the appearance of fortuitous states makes it plausible that they are relevant also at strong coupling in the gravity regime. These ideas have been extended to other holographic setups, such as SYK models~\cite{Chang:2024lxt}, ABJM~\cite{Belin:2025hsg,Behan:2025hbx}, matrix models~\cite{deMelloKoch:2025ngs,deMelloKoch:2025cec,Chen:2025sum}
 and to the D1D5 CFT~\cite{Chang:2025rqy,Hughes:2025tdy,Chang:2025wgo,Zhang:2026jnf}, which is the setup considered here. A key ingredient in these constructions is the definition of a projection operator $\Pi$ that links the CFTs with different values of the central charge: in the case of interest for this paper, $c=6N$ and $\Pi$ project the spectrum of the $N+1$ theory to that of the $N$ theory. 

Monotone states form sequences whose elements are mapped one into the other by $\Pi$ and are characterised by observables, such as three-point couplings, that are smooth functions of $N$ (when interpreted as a continuous parameter). It is natural to assume that graviton gas states are monotone, even if we will see that this is not obvious in sectors where black holes exist. Finding the explicit form of protected states in such sectors is difficult, so it is useful to first look at a cohomology problem defined by the supercharges one wishes to preserve. In the $1/4$-BPS sector of the D1D5 CFT, we identify the differential operator with 
the doublet formed by the right-moving supercharges with positive R-charge, $Q^A \equiv \mathcal{\tilde G}^{+A}_{-1/2}$ ($A$ being an index in the fundamental rep of the $SU(2)$ outer automorphism of the superalgebra). At a first stage one focuses on cohomology classes, given by $Q^A$-closed states up to $Q^A$-exact terms, and only at a second stage one looks for the proper BPS representatives that are also annihilated by $Q^{A\,\dagger} \equiv \mathcal{\tilde G}^{-A}_{1/2}$, the remaining half of the right-moving supercharges. For definiteness, we will explicitly carry out this construction in the sector of states with fixed left-moving dimension and R-symmetry quantum number $(h,j)=(1,0)$, with the goal of identifying monotone and fortuitous states as we change the value of $N$. The analysis is subtle due to fact that the naive projection operator does not commute with the supercharges $Q^A$~\cite{Chang:2025rqy}, a feature shared also by a two flavours SYK model \cite{Chang:2024lxt}.

In order to understand better the properties of monotone and fortuitous states we consider also their three-point couplings with single particles states, i.e. CPOs that correspond to excitations of supergravity fields. These couplings are protected~\cite{Baggio:2012rr} and so we can compare the weak coupling results obtained in conformal perturbation theory against those obtained at strong coupling via holography. Starting from $N=2$, we study how the degeneracy of the free spectrum is lifted in conformal perturbation theory applied to the D1D5 CFT~\cite{Gava:2002xb,Carson:2014yxa,Carson:2014ena,Carson:2015ohj,Carson:2016uwf,Guo:2019pzk,Guo:2019ady,Guo:2020gxm,Guo:2022ifr,Fiset:2022erp,Hughes:2023apl,Hughes:2023fot,Gaberdiel:2023lco,Gaberdiel:2024nge,Gaberdiel:2025smz,Gaberdiel:2026jor}\footnote{In the covering space approach discussed in these references great care is needed to keep track of signs, see for instance the in-depth analysis of~\cite{Burrington:2015mfa}. As discussed in the main text, we will not attempt a first principle derivations of the action of the deformation operator on all states, but use the results for an easier susbsector and reconstruct the full action of the supercharges by imposing consistency conditions coming from the supersymmetry algebra.}. We obtain the explicit form of the candidate fortuitous state found in~\cite{Hughes:2025tdy} and show that, in theories with $N>2$, it becomes part of a long multiplet, consistent with its interpretation as a fortuitous state. We also construct explicit sequences of states that naturally admit an interpretation as monotone, an identification supported by the regular pattern of the three-point couplings mentioned above. In general, the form of such monotone states coincides with their naive graviton-gas representatives only at leading order in the large-$N$ expansion. At finite $N$ the true monotone states contain non-graviton-gas corrections, which are needed to make the states supersymmetric away from the free point. The limit of the monotone sequence for the lowest value of $N$, namely $N=2$, is particularly non-intuitive, as the graviton-gas component disappears in the limit. Despite these non-trivial corrections, we find that the value of the simplest protected three-point coupling containing the monotone state is identical to that computed with the naive graviton-gas form of the state at all orders in $1/N$. This surprising coincidence might explain the success of the precision holography analysis of the superstrata geometries \cite{Giusto:2015dfa,Giusto:2019qig,Rawash:2021pik}, which was based on the naive graviton-gas representation of the dual states.   

Although the monotone nature of the sequence we find is supported by the expected regularity of the three-point couplings, it fails to satisfy the monotone criterion \cite{Chang:2025rqy} based on the naive projection operator $\Pi$. Restoring consistency between the monotone sequence and the projection operator requires a modification\footnote{After this modification $\Pi$ does not satisfy the property of a projector in the standard mathematical sense; for simplicity we will still refer to it as a ``projector''.} of $\Pi$ so that it commutes with the supercharges $Q^A$. We show that this is possible in the sector we consider and provide evidence that it can be done in general.

This paper is organised as follows: In section~\ref{sec:symmo} we briefly review the basics of the symmetric orbifold describing the free regime of the D1D5 CFT. In particular, we highlight the features relevant for the K3 case which has been less studied in the recent literature. We also list the states with $(h,j)=(1,0)$ and $\bar{h}=\bar{j}$ for the $N=2$ and $N=3$ theories. In section~\ref{sec:def} we describe how the right-moving superconformal charges act on the states above, including the first corrections in conformal perturbation theory. We conclude with section~\ref{sec:comm} discussing the lessons that this analysis provides for the general program of studying monotone and fortuitous states in holographic theories. Some technical details are discussed in appendix~\ref{sec:appA}.

\section{The symmetric orbifold Sym$_N$(K3)}
\label{sec:symmo}
The D1D5 CFT dual to type IIB string theory on AdS$_3 \times$ S$^3 \times$ K$3$ has a 84-dimensional superconformal moduli space. The CFT is most easily described at the free point, where the theory is comprised by a collection of 2D free bosons and fermions\footnote{We follow the conventions of~\cite{Hughes:2025tdy} where one can find the explicit expressions of the superalgebra and how it is realised in the free theory.}
\begin{equation}
  \label{eq:elf}
  (\partial X^{A \dot{A}}_{(r)}\;,~ \psi^{\alpha \dot{A}}_{(r)}) \quad   (\bar\partial X^{A \dot{A}}_{(r)}\;,~ \tilde{\psi}^{\dot{\alpha} \dot{A}}_{(r)})\;,
\end{equation}
where $r=1,2,\ldots,N$ is a copy index and the upper indices describe a fundamental representation of various $SU(2)$ groups relevant for the problem. The indices $\alpha$ and $\dot{\alpha}$ transform under the left and right $SU(2)$ R-symmetry groups related to the S$^3$ isometries in the bulk, while $A$ and $\dot{A}$ are $SU(2)$ indices internal to K3:
\begin{align}
\label{eq:su2rep}
\begin{split}
\alpha, \beta = \{+,-\}: & \quad  SU(2)_L,  \qquad  A, B = \{1,2\}: \quad  SU(2)_1, \\[5pt]
\dot\alpha, \dot\beta = \{\dot{+},\dot{-}\}: & \quad  SU(2)_R,  \qquad \dot A, \dot B = \{\dot{1},\dot{2}\}: \quad  SU(2)_2.
\end{split}
\end{align}
The free D1D5 CFT is the symmetric orbifold theory Sym$_N$(K3), composed of multiple copies of a ``seed'' theory CFT on K$3$. We focus on the case where K$3=T^4/Z_2$. There are two orbifolds relevant for the Sym$_N(T^4/Z_2)$ theory: the one of the symmetric group $S_N$, permuting the copy indices $(r)$, and the orbifold of $Z_2$ which multiplies by $-1$ every field that carries one index $\dot{A}$. As pointed out in~\cite{Baggio:2015jxa}, these operations combine into the semidirect product of the two groups and we need to consider the symmetric orbifold
\begin{equation}
  \label{eq:semdpr}
  {\rm Sym}_N(T^4/Z_2) = (T^4)^N/(S_N \ltimes Z_2^N)\;.
\end{equation}
The Hilbert space of the Sym$_N(T^4/Z_2)$ theory can be decomposed into different twisted sectors, consisting of multiple copies of the seed CFT joined together. The Hilbert space of the CFT resulting from joining $k$ seed CFTs contains states, or ``strands'', of winding $k$.

Each element of $S_N \ltimes Z_2^N$ is a collection of $N$ $\pm$ signs and a permutation of $S_N$. The elements act on the elementary fields~\eqref{eq:elf} by first multiplying each of them by the corresponding sign and then applying the permutation on the copy indices $(r)$. As an example, the element $(1_-, 2_-)$ in the $N=2$ theory implements the following monodromies
\begin{equation}
  \label{eq:1m2m}
  \begin{pmatrix} \partial X^{A \dot{A}}_{(1)} \\ \partial X^{A \dot{A}}_{(2)} \end{pmatrix}_{\!\!e^{2\pi i}z}
  =  \begin{pmatrix} 0 & -1 \\ -1 & 0 \end{pmatrix} \begin{pmatrix} \partial X^{A \dot{A}}_{(1)} \\ \partial X^{A \dot{A}}_{(2)} \end{pmatrix}_{\!\!z}
\end{equation}
and similarly for the fermions. As usual, the twisted sectors are in one-to-one correspondence with the conjugagy classes of the group and, for our purposes, the elements with a single length-two cycle are important. Still focusing on $N=2$ for simplicity, there are two conjugacy classes: one containing the elements $(1_+, 2_+)$ and $(1_-, 2_-)$, and the other the elements $(1_+, 2_-)$ and $(1_+, 2_-)$. Only the first conjugacy class contains a CPO, while the second contains only non-BPS states, see~\cite{Baggio:2015jxa} for more details.

In a twist sector, the elementary fields are periodic up to a transformation in the corresponding conjugacy class, so it is convenient to introduce the basis diagonalising the monodromies. For instance, for the elements $(1_+, 2_+)$ and $(1_-, 2_-)$, we have
\begin{equation}
  \label{eq:r2rho}
  \partial X^{A \dot{A}}_{\rho=0} = \frac{1}{\sqrt{2}} \left[ \partial X^{A \dot{A}}_{(1)} + \partial X^{A \dot{A}}_{(2)}\right]\quad,\quad \partial X^{A \dot{A}}_{\rho=1} = \frac{1}{\sqrt{2}} \left[ \partial X^{A \dot{A}}_{(1)} - \partial X^{A \dot{A}}_{(2)}\right]\;.
\end{equation}
The $\rho=0$ ($\rho=1$) fields are periodic (antiperiodic) for $(1_+, 2_+)$ and the role is swapped for $(1_-, 2_-)$. Thus we can describe the CPO in this sector
\begin{equation}
  \label{eq:Sigmapp}
  \Sigma^{++}_2= \mathds{1}_{\rho=0} \otimes \left(\sigma_{\rho=1} \,e^{\frac{i}{2}(H+K)_{\rho=1}}\right) \left(\tilde \sigma_{\rho=1} \,e^{\frac{i}{2}(\tilde{H}+\tilde{K})_{\rho=1}}\right) + (\rho=0 \leftrightarrow \rho=1)\;,
\end{equation}
where $\sigma$ are the bosonic twist fields changing the monodromies of $\partial X^{A \dot{A}}_\rho$ to antiperiodic and the exponentials are the bosonised description of the spin fields that implement the same change on the fermions. The bosonic fields $H$, $K$ are defined via the bosonization relations 
  \begin{equation}
      \label{eq:ermbosdef}
  \begin{aligned} 
 \psi^{+\dot{1}}_{(\rho)} &= i \,  e^{iK_{(\rho)} } , 
  &\psi^{-\dot{2}}_{(\rho)} &= i \, e^{-iK_{(\rho)}}  \,,& \\ \psi^{+\dot{2}}_{(\rho)} &=  e^{ iH_{(\rho)}}  , 
  &\psi^{-\dot{1}}_{(\rho)} &=  e^{-iH_{(\rho)}}\ ,&
 \end{aligned}
 \end{equation}
with analogous right-moving expressions. The writing above is imprecise as cocycle factors should be included in order to reproduce the correct anticommutation relations between different fermions: we keep track of these signs by hand and refer to~\cite{Burrington:2015mfa} for a systematic approach. The symmetrisation between $\rho=0$ and $\rho=1$ in \eqref{eq:Sigmapp} reflects the presence of two elements in the same conjugacy class and is a novelty of the $T^4/Z_2$ case. The operator~\eqref{eq:Sigmapp} has $h=j=1/2$ and similarly in the antiholomorphic sector: its $h=1$, $j=0$ superdescendants describe the superconformal deformations of the symmetric orbifold~\eqref{eq:semdpr} that we will consider in the next section.

The free fermions and bosons can be decomposed into modes via 
\begin{equation}
\partial X^{A \dot A}(z)
  = \sum_{n} \alpha^{A \dot A}_{n}\, z^{-n-1}\quad,
\quad
\psi^{\alpha \dot A} (z)
  = \sum_{r} \psi^{\alpha \dot A}_{r}\, z^{-r-\frac12},
\end{equation}
with similar expressions holding for the antiholomorphic fields $\bar\partial X^{A \dot{A}}_{(r)}$ and $\tilde{\psi}^{\dot{\alpha} \dot{A}}_{(r)}$.  The allowed mode values depend on the length $k$ of the state and whether the state is $Z_2$-twisted or not. The allowed free bosonic modes are
\begin{equation}
    Z_2\text{-untwisted}: \quad n=\frac{m}{k}, \qquad Z_2\text{-twisted}: \quad n = \frac{m}{k}+ \frac{1}{2k}\,,
\end{equation}
where $m \in \mathbb{Z}$. The allowed modes $r$ for the free NS fermion $\psi^{\alpha \dot A}_{r}$ on a length-$k$ strand are 
\begin{equation}
    Z_2\text{-untwisted}: \quad r=\frac{m}{k}+\frac{1}{2}, \qquad Z_2\text{-twisted}: \quad  r=\frac{m}{k}- \frac{1}{2k}+\frac{1}{2}.
\end{equation}
The superalgebra associated with the D1D5 CFT on ${\rm Sym}_N$(K3) is the small $\mathcal{N}=(4,4)$ superconformal algebra with central charge $c = 6N$. The symmetry algebra comprises a left-moving stress tensor $T$, $SU(2)_L$ R-symmetry currents $J^a$, where $a =1,2,3$, and supersymmetry generators $G^{\alpha A}$, each with right-moving counterparts $\tilde{T}$, $\tilde{J}^a$, and $\tilde{G}^{\dot{\alpha} A}$. The currents can be expanded in terms of modes $L_n$, $J^a_n$ and $G^{\alpha A}_r$, which can be expressed as sums of the free fermion and boson modes
\begin{align} \label{genmodes}
\begin{split}
    L_n &= \frac{1}{2} \varepsilon_{AB}\, \varepsilon_{\dot A \dot B}\, \sum_{m } :\, \alpha^{A\dot{A}}_{m} \alpha^{B\dot{B}}_{n-m} \, : + \, \frac{1}{2} \varepsilon_{\alpha \beta}\, \varepsilon_{\dot A \dot B}\, \sum_{r} \left( -n +m -\frac{1}{2}\right):\, \psi^{\alpha \dot{A}}_{r} \psi^{\beta \dot{B}}_{n-r} :  \\
    G^{\alpha A}_r &= \, \varepsilon_{\dot A \dot B}\, \sum_{m } :\, \alpha^{A\dot{A}}_{m} \psi^{\alpha\dot{B}}_{r-m} \, : \\
    J^a_n &=  \frac{1}{2} \varepsilon_{\alpha \beta} \varepsilon_{\dot A \dot B}\, (\sigma^{aT})^{\beta}{}_{\gamma}\,\sum_{m } :\, \psi^{\alpha\dot{A}}_{m} \psi^{\gamma\dot{B}}_{n-m} \, :. \\
\end{split}
\end{align}
The matrices $({\sigma^{a})^{\alpha}}_{\beta}$ are the usual Pauli matrices. It is useful to also define the ``global'' anomaly-free subalgebra generated by the modes $L_0$, $L_{\pm 1}$, $J^a_0$, $G^{\alpha A}_{\pm \frac{1}{2}}$. 

We are interested in constructing 1/4-BPS states within the CFT by dressing 1/2-BPS chiral-chiral primaries with left-moving excitations. The chiral-chiral primaries in the NS sector in the length-$k$ sector are denoted as
\begin{equation}
\label{chiralprim}
\begin{aligned}
|\alpha \dot{\alpha}\rangle_k
    &:\qquad h=j=\frac{k+\alpha}{2},
    & \qquad \tilde h=\tilde j=\frac{k+\dot \alpha}{2},
    \\
|\dot A\dot B\rangle_k
    &:\qquad h=j=\frac{k}{2},
    & \tilde h=\tilde j=\frac{k}{2} \quad \ \  ,
    \\
|r\rangle_k
    &:\qquad h=j=\frac{k}{2},
    & \tilde h=\tilde j=\frac{k}{2} \quad \ \ ,
\end{aligned}
\end{equation}
where $\alpha,\dot{\alpha} = \pm 1$. The 16 states $|r\rangle_k $ belong to the $Z_2$-twisted sector. The $Z_2$-untwisted chiral primaries can be written as fermionic modes acting on the length-$k$ vacuum $ |-\dot{-}\rangle_k$
\begin{align}
\begin{split}
   & |-\dot{+}\rangle_k  = \tilde{\psi}^{\dot{+} \dot{1}}_{-\frac{1}{2}} \tilde{\psi}^{\dot{+} \dot{2}}_{-\frac{1}{2}}  \, |-\dot{-}\rangle_k\qquad\qquad\,\,\,\,,\quad |+\dot{-}\rangle_k  = {\psi}^{{+} \dot{1}}_{-\frac{1}{2}} {\psi}^{{+} \dot{2}}_{-\frac{1}{2}}  \, |-\dot{-}\rangle_k\;, \\
   & |+\dot{+}\rangle_k  = {\psi}^{{+} \dot{1}}_{-\frac{1}{2}} {\psi}^{{+} \dot{2}}_{-\frac{1}{2}} \,\tilde{\psi}^{\dot{+} \dot{1}}_{-\frac{1}{2}} \tilde{\psi}^{\dot{+} \dot{2}}_{-\frac{1}{2}}  \,  |-\dot{-}\rangle_k\,\quad,\quad |\dot{A}\dot{B}\rangle_k  = {\psi}^{+ \dot{A}}_{-\frac{1}{2}} \tilde{\psi}^{\dot{+} \dot{B}}_{-\frac{1}{2}}  \, |-\dot{-}\rangle_k \;.
\end{split}
\end{align}
As one moves away from the free orbifold point to a generic point in moduli space, states may ``lift'', i.e. recombine and form a long supersymmetry multiplet. In doing so, states gain an anomalous dimension and no longer preserve the same supersymmetries compared to the free point. 

The CFT elliptic genus, or CFT index, provides a signed count of 1/4-BPS states for a given value of $N$, which holds at all points in the moduli space of the CFT \cite{Hughes:2025tdy}. For the NS sector of the D1D5 CFT on ${\rm Sym}_N$(K3), the generating function of elliptic genera $\mathcal{E}^{\mathrm{CFT}}_{\mathrm{NS}}(q,y;N)$ takes the form
\begin{equation}
Z^{\mathrm{CFT}}_{\mathrm{NS},K3}(p,q,y)
=
\sum_{N=1}^{\infty}
p^N \,
\mathcal{E}^{\mathrm{CFT}}_{\mathrm{NS}}(q,y;N)
=
\prod_{n,m,\ell}
\left(
1 - p^n q^m y^\ell
\right)^{
-c(4 n m - n^2 - \ell^2)
},
\end{equation}
where $n \in \mathbb{Z}_{\ge 1}$, $
m \in \mathbb{Z}_{\ge 0}$ and $
\ell \in \mathbb{Z}$. The fugacities $p,q,y$ provide a count of states for a given set of left-moving quantum numbers. The coefficients \(c(n)\) are a property of the seed sigma model on K$3$ and can be defined in terms of Jacobi theta functions, more detail can be found in \cite{Dijkgraaf:1996xw,Hughes:2025tdy}. A supergraviton elliptic genus $\mathcal{E}^{\mathrm{graviton}}_{\mathrm{K3}}(q,y;N)$ can also be defined by restricting the Hilbert space of the CFT to that of the supergravitons \cite{deBoer:1998us}. These are states whose strands are chiral-chiral primaries or descendants thereof obtained by acting only with the global subalgebra. 
The $q$-expansions for the CFT and supergraviton indices at $N=2$ are 
\begin{align}\label{eq:genusn2}
\begin{split}
\mathcal{E}^{\mathrm{CFT}}_{\mathrm{NS}}(q,y;2)
&\approx
3
+\sqrt{q}\,\left(42y^{-1}+42y\right)
+q\left(234y^{-2}+360+234y^{2}\right)+\cdots \;,
\\[1ex]
\mathcal{E}^{\mathrm{graviton}}_{K3}(q,y;2)
&\approx
3
+\sqrt{q}\,\left(42y^{-1}+42y\right)
+q\left(234y^{-2}+340+234y^{2}\right)
+\cdots \;.
\end{split}
\end{align}
States contributing to the index at order $q^{h} y^{2j}$ correspond to chiral-chiral primaries with left-moving excitations, giving the $(L_0,J^3_0)$ quantum numbers $(h,j)$.  For $N=2$, the CFT and supergraviton indices begin to differ at $(h,j)=(1,0)$. As shown in \cite{Hughes:2025tdy}, the difference between the two indices is due to suspected fortuitous states rather than singletons, i.e. states obtained by acting on a full chiral primary -- rather than on individual strands -- with any element of the superlagebra. At $N=3$, the CFT and supergraviton elliptic genera agree up to order $(h,j)=(1,0)$:
\begin{align}
\begin{split}
\mathcal{E}^{\mathrm{CFT}}_{\mathrm{NS}}(q,y;3)
&\approx
4
+\sqrt{q}\,\left(64y^{-1}+64y\right)
+q\left(508y^{-2}+800+508y^{2}\right)+\cdots 
\\[1ex]
\mathcal{E}^{\mathrm{graviton}}_{K3}(q,y;3)
&\approx
4
+\sqrt{q}\,\left(64y^{-1}+64y\right)
+q\left(508y^{-2}+800+508y^{2}\right)
+\cdots \,.
\end{split}
\end{align}
Table \ref{tab:N2h1j0} and Table \ref{tab:N3h1j0} show the $1/4$-BPS states  with left-moving quantum numbers $(h,j)=(1,0)$, for $N=2$ and $N=3$ respectively. The tables show a colour-coded count of states, where the supergravitons are shown in red and the non-supergravitons are shown in blue. It would be interesting to generalise this analysis by using the K3 version of the refined index discussed in~\cite{Hughes:2025oxu,Hughes:2026qqn}, as this should allow to isolate smaller sectors and so to increase the values of $h$.

\begin{table}[ht]
\centering
\caption{$|h,j\rangle = |1, 0\rangle$ sector for $N=2$. 360 CFT states and 340 supergravitons.}
\renewcommand{\arraystretch}{1.3}
\resizebox{0.65\textwidth}{!}{%
\begin{tabular}{|c|c|c|c|c|}
\hline
{State}   & {$(\tilde{j}, \mathbf{d_1},\mathbf{d_2},\mathbf{\tilde{d}_2} )_s$} & {Count} \\
\hline

$J^{-}_{0} |\dot{A}\dot{B} \rangle_1  \otimes |\dot{C}\dot{D} \rangle_1$ & {{$(1, \mathbf{1},\mathbf{1},\mathbf{1} )_1+(1, \mathbf{1},\mathbf{3},\mathbf{1} )_1+(1, \mathbf{1},\mathbf{1},\mathbf{3}  )_1 + (1, \mathbf{1},\mathbf{3},\mathbf{3}  )_1$}} &\textcolor{red}{16} \\[4pt]

 $J^{-}_{0} |\dot{A}\dot{B} \rangle_1 \otimes |r \rangle_1$  & {$(1, \mathbf{1},\mathbf{2},\mathbf{2} )_{1r}$} & \textcolor{red}{64} \\[4pt]

 $J^{-}_{0} |r \rangle_1 \otimes |\dot{A}\dot{B} \rangle_1$ & {$(1, \mathbf{1},\mathbf{2},\mathbf{2} )_{1r}$} &\textcolor{red}{64} \\[4pt]

 $J^{-}_{0} |r \rangle_1 \otimes |r \rangle_1$ &  {$(1, \mathbf{1},\mathbf{1},\mathbf{1} )_{1rr}$}& \textcolor{red}{256} \\[4pt]

\hline

$  G^{-A}_{-\frac{1}{2}} |r \rangle_1 \otimes |-\dot{\alpha} \rangle_1$ & {$(\frac{1}{2}, \mathbf{2},\mathbf{1},\mathbf{1} )_{1r}+(\frac{3}{2}, \mathbf{2},\mathbf{1},\mathbf{1})_{1r}$}  & $\textcolor{red}{-64}$ \\[4pt]

 $  G^{-A}_{-\frac{1}{2}} |\dot{A} \dot{B}\rangle_1 \otimes |-\dot{\alpha} \rangle_1$  & {$(\frac{1}{2}, \mathbf{2},\mathbf{2},\mathbf{2} )_1+(\frac{3}{2}, \mathbf{2},\mathbf{2},\mathbf{2})_1$} & $\textcolor{red}{-16}$ \\[4pt]

 $  J^{-}_{0} |+ \dot{\alpha}\rangle_1 \otimes |-\dot{\beta} \rangle_1$  & {$(0, \mathbf{1},\mathbf{1},\mathbf{1} )_1+2(1, \mathbf{1},\mathbf{1},\mathbf{1}  )_1 + (2, \mathbf{1},\mathbf{1},\mathbf{1}  )_1$}  & $\textcolor{red}{4}$ \\[4pt]

$|- \dot{\alpha}\rangle_1 \otimes    \tau^i_{\dot{A}\dot{B}}\alpha^{A\dot B}_{-1/2} \psi^{-\dot A}_{0} |r\rangle_1$ &  {$(\frac{1}{2}, \mathbf{2},\mathbf{3},\mathbf{1} )_{1r}+(\frac{3}{2}, \mathbf{2},\mathbf{3},\mathbf{1})_{1r}$} & $\textcolor{blue}{-192}$ \\[4pt]

 $  \tau^i_{\dot{A}\dot{B}}\psi^{-\dot A}_{-1/2}\psi^{+\dot B}_{-1/2} |- \dot{\alpha}\rangle_1 \otimes |- \dot{\beta}\rangle_1$  & \ $(0, \mathbf{1},\mathbf{3},\mathbf{1} )_1+2(1, \mathbf{1},\mathbf{3},\mathbf{1}  )_1 + (2, \mathbf{1},\mathbf{3},\mathbf{1}  )_1$ & $\textcolor{blue}{12}$ \\[4pt]
\hline

$G^{-A}_{-\frac{1}{2}} |- \dot{\alpha}\rangle_2$  & {$(\frac{1}{2}, \mathbf{2},\mathbf{1},\mathbf{1}  )_2 + (\frac{3}{2}, \mathbf{2},\mathbf{1},\mathbf{1}  )_2$}  & $\textcolor{red}{-4}$  \\[4pt]
 $ J^-_0|\dot{A}\dot{B} \rangle_2$ & {$(1, \mathbf{1},\mathbf{2},\mathbf{2}   )_2$}  & $\textcolor{red}{4}$ \\[4pt]
$J_0^-|r\rangle_2$& {$(1, \mathbf{1},\mathbf{1},\mathbf{1} )_{2r}$} & $\textcolor{red}{16}$ \\[4pt]

 $L_{-\frac{1}{2}} J^-_{\frac{1}{2}} |r\rangle_2$ & {$(1, \mathbf{1},\mathbf{1},\mathbf{1} )_{2r}$} & $\textcolor{blue}{16}$ \\[4pt]

  $L_{-\frac{1}{2}} J^-_{\frac{1}{2}} |\dot{A}\dot{B}\rangle_2$ & {$(1, \mathbf{1},\mathbf{2},\mathbf{2} )_{2}$} & $\textcolor{blue}{4}$ \\[4pt]

  $\tau^i_{\dot{A}\dot{B}}\alpha^{A\dot A}_{-1/2} \psi^{-\dot B}_{0} |- \dot{\alpha}\rangle_2$  &  {$(\frac{1}{2}, \mathbf{2},\mathbf{3},\mathbf{1}  )_2 + (\frac{3}{2}, \mathbf{2},\mathbf{3},\mathbf{1}  )_2$} & $\textcolor{blue}{-12}$  \\[4pt]

  $\tau^i_{\dot{C}\dot{D}} \tau^j_{CD} \alpha^{C\dot C}_{-1/4}\alpha^{D\dot D}_{-1/4}
   \epsilon_{\dot{A}\dot{B}}\psi^{-\dot A}_{1/4}\psi^{-\dot B}_{1/4} |r\rangle_2$& {$(1, \mathbf{3},\mathbf{3},\mathbf{1} )_{2r}$} & $\textcolor{blue}{144}$ \\[4pt]

 $\tau^i_{\dot{A}\dot{B}}\psi^{-\dot A}_{-1/4}
   \psi^{-\dot{B}}_{1/4} |r\rangle_2$ & {$(1, \mathbf{1},\mathbf{3},\mathbf{1} )_{2r}$} & $\textcolor{blue}{48}$ \\[4pt]
   
\hline
\end{tabular}
}
\label{tab:N2h1j0}
\end{table}

\begin{table}[ht]
\centering
\caption{$|h,j\rangle = |1, 0\rangle$ sector for $N=3$. 800 CFT states and 800 supergravitons.}
\resizebox{0.75\textwidth}{!}{%
\renewcommand{\arraystretch}{1.3}
\begin{tabular}{|c|c|c|c|c|c|}
\hline
State  & {$(\tilde{j}, \mathbf{d_1},\mathbf{d_2}, \mathbf{\tilde{d}_2} )_s$} & Count \\
\hline

$  |-\dot{\alpha}\rangle_1 \otimes |\dot{A}\dot{B}\rangle_1 \otimes J^{-}_{0}|r\rangle_1  $
& {$(1, \mathbf{1},\mathbf{2}, \mathbf{2} )_1+(2, \mathbf{1},\mathbf{2}, \mathbf{2} )_1$} & $\textcolor{red}{128 }$ \\[6pt] 
 $  |-\dot{\alpha}\rangle_1 \otimes |r\rangle_1 \otimes J^{-}_{0}|r\rangle_1  $
& {$(1, \mathbf{1},\mathbf{1}, \mathbf{1} )_1+(2, \mathbf{1},\mathbf{1}, \mathbf{1} )_1$} & $\textcolor{red}{512 }$ \\[6pt] 
$  J^{-}_{0}|\dot{A}\dot{B}\rangle_1 \otimes |-\dot{\alpha}\rangle_1 \otimes |r\rangle_1  $
& {$(1, \mathbf{1},\mathbf{2}, \mathbf{2} )_1+(2, \mathbf{1},\mathbf{2}, \mathbf{2} )_1$} & $\textcolor{red}{128}$ \\[6pt] 

$  J^{-}_{0}|\dot{A}\dot{B}\rangle_1 \otimes |\dot{A}\dot{B}\rangle_1 \otimes |-\dot{\alpha}\rangle_1  $
&  {$(1, \mathbf{1},\mathbf{1}, \mathbf{1} )_1+(1, \mathbf{1},\mathbf{3}, \mathbf{1} )_1+(1, \mathbf{1},\mathbf{1}, \mathbf{3} )_1+(1, \mathbf{1},\mathbf{3}, \mathbf{3} )_1+ (2, \mathbf{1},\mathbf{1}, \mathbf{1} )_1+(2, \mathbf{1},\mathbf{3}, \mathbf{1} )_1+(2, \mathbf{1},\mathbf{1}, \mathbf{3} )_1+(2, \mathbf{1},\mathbf{3}, \mathbf{3} )_1$} & $\textcolor{red}{32}$ \\[6pt] 

\hline

$  G^{-A}_{-1/2}|\dot{A}\dot{B}\rangle_1 \otimes |-\dot{\beta}\rangle_1 \otimes |-\dot{\gamma}\rangle_1  $
& {$(\tfrac{1}{2}, \mathbf{2},\mathbf{2}, \mathbf{1} )_{1} +2(\tfrac{3}{2}, \mathbf{2},\mathbf{2}, \mathbf{1} )_{1} +(\tfrac{5}{2}, \mathbf{2},\mathbf{2}, \mathbf{1} )_{1}$} & $\textcolor{red}{-24}$ \\[6pt] 

         $  G^{-A}_{-1/2}|r\rangle_1 \otimes |-\dot{\beta}\rangle_1 \otimes |-\dot{\gamma}\rangle_1 $  & {$(\tfrac{1}{2}, \mathbf{2},\mathbf{1}, \mathbf{1} )_{1r} +(\tfrac{3}{2}, \mathbf{2},\mathbf{1}, \mathbf{1} )_{1r} +(\tfrac{5}{2}, \mathbf{2},\mathbf{1}, \mathbf{1} )_{1r}  $} & $\textcolor{red}{-96}$  \\[6pt] 
$  J^{-}_{0}|+\dot{\alpha}\rangle_1 \otimes |-\dot{\beta}\rangle_1 \otimes |-\dot{\gamma}\rangle_1  $
& {$(0, \mathbf{1},\mathbf{1}, \mathbf{1} )_1+2(1, \mathbf{1},\mathbf{1}, \mathbf{1} )_1 +2(2, \mathbf{1},\mathbf{1}, \mathbf{1} )_1 +(3, \mathbf{1},\mathbf{1}, \mathbf{1} )_1 $}  & $\textcolor{red}{6}$ \\[6pt]

 $  \tau^i_{\dot{A}\dot{B}}\, \alpha^{A\dot{A}}_{-1/2} \, \psi^{-\dot{B}}_{0} |r\rangle_1 \otimes |-\dot{\beta}\rangle_1 \otimes |-\dot{\gamma}\rangle_1 $ &  $(\tfrac{1}{2}, \mathbf{2},\mathbf{3}, \mathbf{1} )_{1r} +(\tfrac{3}{2}, \mathbf{2},\mathbf{3}, \mathbf{1} )_{1r} +(\tfrac{5}{2}, \mathbf{2},\mathbf{3}, \mathbf{1} )_{1r}$  & $\textcolor{blue}{-288}$  \\[6pt] 

$\tau^i_{\dot{A}\dot{B}}\, \psi^{-\dot{A}}_{-1/2} \, \psi^{+\dot{B}}_{-1/2} |- \dot{\alpha}\rangle_1  \otimes|-\dot{\beta}\rangle_1 \otimes |-\dot{\gamma}\rangle_1 $  & $(0, \mathbf{1},\mathbf{3}, \mathbf{1} )_1+2(1, \mathbf{1},\mathbf{3}, \mathbf{1} )_1 +2(2, \mathbf{1},\mathbf{3}, \mathbf{1} )_1 +(3, \mathbf{1},\mathbf{3}, \mathbf{1} )_1$ & $\textcolor{blue}{18}$ \\[6pt]

\hline

         $  |\dot{A}\dot{B}\rangle_1 \otimes J^{-}_{0}|-\dot{\alpha}\rangle_2  $
&{$(1, \mathbf{1},\mathbf{2}, \mathbf{2} )_1+(2, \mathbf{1},\mathbf{2}, \mathbf{2} )_1$} & $\textcolor{red}{8 }$ \\[6pt] 
  
         $  J^{-}_{0}|-\dot{\alpha}\rangle_2 \otimes |r\rangle_1  $
&  {$(1, \mathbf{1},\mathbf{1}, \mathbf{1} )_2+(2, \mathbf{1},\mathbf{1}, \mathbf{1} )_2$} & $\textcolor{red}{32 }$ \\[6pt] 

\hline

         $  J^{-}_{0}|\dot{A}\dot{B}\rangle_1 \otimes |-\dot{\alpha}\rangle_2  $
& {$(1, \mathbf{1},\mathbf{2}, \mathbf{2} )_2+(2, \mathbf{1},\mathbf{2}, \mathbf{2} )_2$} & $\textcolor{red}{8}$ \\[6pt] 

         $  |-\dot{\alpha}\rangle_2 \otimes J^{-}_{0}|r\rangle_1  $
&  {$(1, \mathbf{1},\mathbf{1}, \mathbf{1} )_2+(2, \mathbf{1},\mathbf{1}, \mathbf{1} )_2$} & $\textcolor{red}{32 }$ \\[6pt] 

\hline

        $  |-\dot{\alpha}\rangle_1 \otimes J^{-}_{0}|\dot{A}\dot{B}\rangle_2  $
&  {$(1, \mathbf{1},\mathbf{2}, \mathbf{2} )_2+(2, \mathbf{1},\mathbf{2}, \mathbf{2} )_2 $}  & $\textcolor{red}{8 }$ \\[6pt]

         $  |-\dot{\alpha}\rangle_1 \otimes J^{-}_{0}|r\rangle_2  $
 & {$(1, \mathbf{1},\mathbf{1}, \mathbf{1} )_{2r} +(2, \mathbf{1},\mathbf{1}, \mathbf{1} )_{2r} $} & $\textcolor{red}{32}$ \\[6pt]

$  G^{-A}_{-1/2}|-\dot{\alpha}\rangle_2 \otimes |-\dot{\beta}\rangle_1  $
 &  {$(\tfrac{1}{2}, \mathbf{2},\mathbf{1}, \mathbf{1} )_{2} +2(\tfrac{3}{2}, \mathbf{2},\mathbf{1}, \mathbf{1} )_{2}+(\tfrac{5}{2}, \mathbf{2},\mathbf{1}, \mathbf{1} )_{2} \ \ $}  & $\textcolor{red}{-8}$ \\[6pt] 
 
{$  |-\dot{\alpha}\rangle_1 \otimes L_{-\frac{1}{2}} J^{-}_{\frac{1}{2}}|\dot{A}\dot{B}\rangle_2  $}
& {$(1, \mathbf{1},\mathbf{2}, \mathbf{2} )_2+(2, \mathbf{1},\mathbf{2}, \mathbf{2} )_2 $}  & $\textcolor{blue}{8 }$ \\[6pt]

         $  |-\dot{\alpha}\rangle_1 \otimes L_{-\frac{1}{2}} J^{-}_{\frac{1}{2}}|r\rangle_2  $
 & {$(1, \mathbf{1},\mathbf{1}, \mathbf{1} )_{2r} +(2, \mathbf{1},\mathbf{1}, \mathbf{1} )_{2r} $} & $\textcolor{blue}{32}$ \\[6pt] 

 $|-\rangle_1  \otimes \tau^i_{\dot{A}\dot{B}}\, \psi^{-\dot{A}}_{-1/4} \,  \psi^{-\dot{B}}_{1/4} \, |r\rangle_2 $ &  $(1, \mathbf{1},\mathbf{3}, \mathbf{1} )_{2r} +(2, \mathbf{1},\mathbf{3}, \mathbf{1} )_{2r}$ & $\textcolor{blue}{96}$ \\[6pt]

 $\tau^i_{AB} \tau^j_{\dot{A}\dot{B}}\alpha^{A\dot{A}}_{-1/4} \, \alpha^{B\dot{B}}_{-1/4} \, \epsilon_{\dot{C}\dot{D}} \psi^{-\dot{C}}_{1/4} \, \psi^{-\dot{D}}_{1/4} |r\rangle_2  \otimes |- \dot{\alpha}\rangle_1 $ & $(1, \mathbf{3},\mathbf{3}, \mathbf{1} )_{2r} +(2, \mathbf{3},\mathbf{3}, \mathbf{1} )_{2r}$ & $\textcolor{blue}{288}$ \\[6pt] 

 $  \tau^i_{\dot{A}\dot{B}}\, \alpha^{A\dot{A}}_{-1/2} \, \psi^{-\dot{B}}_{0}|-\dot{\alpha}\rangle_2 \otimes |-\dot{\beta}\rangle_1  $ &   $(\tfrac{1}{2}, \mathbf{2},\mathbf{3}, \mathbf{1} )_{2} +2(\tfrac{3}{2}, \mathbf{2},\mathbf{3}, \mathbf{1} )_{2}+(\tfrac{5}{2}, \mathbf{2},\mathbf{3}, \mathbf{1} )_{2}$ & $\textcolor{blue}{-24}$ \\[6pt]

\hline

$J_0^- |- \dot{\alpha}\rangle_3 $ & {$(1, \mathbf{1},\mathbf{1}, \mathbf{1} )_{3}+(2, \mathbf{1},\mathbf{1}, \mathbf{1} )_{3} $} & $\textcolor{red}{2}$ \\[6pt] 

$G^{-A}_{-\frac{1}{6}} J^-_{\frac{2}{3}} |\dot{A}\dot{B}\rangle_3$ & $(\tfrac{3}{2}, \mathbf{2},\mathbf{2}, \mathbf{2} )_{3}$ & $\textcolor{blue}{-8}$ \\[6pt]

$G^{-A}_{-\frac{1}{6}} J^-_{\frac{2}{3}} |r\rangle_3$  & $(\tfrac{3}{2}, \mathbf{2},\mathbf{1}, \mathbf{1} )_{3r}+(\tfrac{3}{2}, \mathbf{2},\mathbf{1}, \mathbf{1} )_{3r}$ & $\textcolor{blue}{-32}$ \\[6pt]

$\tau^i_{\dot{A}\dot{B}}\,\psi^{-\dot{A}}_{-1/6} \, \psi^{-\dot{B}}_{1/6} \, |- \dot{\alpha}\rangle_3 $ & $(1, \mathbf{1},\mathbf{3}, \mathbf{1} )_{3}+(2, \mathbf{1},\mathbf{3}, \mathbf{1} )_{3}$ & $\textcolor{blue}{6}$ \\[6pt]

$\tau^i_{\dot{A}\dot{B}} \alpha^{A\dot{A}}_{-1/6} \, \psi^{-\dot{B}}_{0} \, J^{-}_{\frac{2}{3}}\,  |r\rangle_3  $ & $(\tfrac{1}{2}, \mathbf{2},\mathbf{3}, \mathbf{1} )_{3r}+(\tfrac{3}{2}, \mathbf{2},\mathbf{3}, \mathbf{1} )_{3r}$ & $\textcolor{blue}{-96}$ \\[6pt]

\hline
\end{tabular}
}
\label{tab:N3h1j0}
\end{table}

All states should be considered as an $S_N \ltimes Z_2^N$-invariant sum of terms. The precise form of lifted states is complicated, but the behavior of states under $SU(2)$ transformations (\ref{eq:su2rep}) is preserved under the twist two deformation~\eqref{eq:margdef} off the free point. It is therefore useful to label states according to their various $SU(2)$ quantum numbers. In particular, states are labelled by $(\tilde{j}, \mathbf{d_1},\mathbf{d_2}, \mathbf{\tilde{d}_2} )_s$, where $\tilde{j}$ is the $SU(2)_R$ R-charge,  $\mathbf{d_1}$ is the $SU(2)_1$ representation, and $\mathbf{d_2}$ and $\mathbf{\tilde{d}_2}$ are the $SU(2)_2$ representations of the left- and right-moving state respectively. The subscript $s$ denotes the strand structure of the state, and whether it is $Z_2$-twisted. The $(\tilde{j}, \mathbf{d_1},\mathbf{d_2}, \mathbf{\tilde{d}_2} )_s$ labels are shown for each state in Table \ref{tab:N2h1j0} and Table \ref{tab:N3h1j0}. In those tables we use the symbol $\tau^i_{\dot A \dot B}$ ($i=1,2,3$) to denote a basis of the rep $\mathbf{3}$ of  $SU(2)_2$, see~\ref{eq:paulimat}.

We are interested in deforming the CFT away from the symmetric orbifold point towards the gravity regime. This corresponds to moving in a particular direction of moduli space, which involves deforming the theory using superdescendants of the twist-2 operator $\Sigma_2^{\alpha \dot{\alpha}}$. In particular, we deform the action through
\begin{equation} \label{eq:margdef}
    S \rightarrow S + \frac{\lambda}{4} \int d^2 z \, \varepsilon_{A B} \varepsilon_{\alpha \beta} \varepsilon_{\dot{\alpha} \dot{\beta}} \oint_z dw\,G^{\alpha A}(w) \oint_{\bar z} d\bar{w}\,{\tilde G}^{\beta B}(\bar{w})\, \Sigma^{\beta \dot{\beta}}_2(z,\bar{z}).
\end{equation}
Under such a deformation, the left- and right-moving supercharges also deform away from their free forms $G^{
{\alpha} A}$ and $\tilde{G}^{
\dot{\alpha} A}$, e.g.
\begin{equation} \label{eq:deformG}
    \tilde{\mathcal{G}}^{
\dot{\alpha} A} = \tilde{G}^{
\dot{\alpha} A} + \lambda\,\delta \tilde{\mathcal{G}}^{
\dot{\alpha} A}
\end{equation}
at first order. 

All 1/4-BPS states with $\tilde h = \tilde j$ are annihilated by the free supercharges ${\tilde G}^{\pm A}_{\mp 1/2}$, but the action of $\delta \mathcal{\tilde G}^{\pm A}_{\mp 1/2}$ can be non-trivial. 
Lifted states are no longer annihilated by the deformed supercharge and combine to form long supermultiplets (quartets)
\begin{align} \label{eq:quartet}
    &(\tilde{j}, \mathbf{d_1},\mathbf{d_2}, \mathbf{\tilde{d}_2} )_{s_1}+(\tilde{j}+\tfrac{1}{2}, \mathbf{d_1}\otimes \mathbf{2},\mathbf{d_2}, \mathbf{\tilde{d}_2} )_{s_2}+(\tilde{j}+1, \mathbf{d_1},\mathbf{d_2}, \mathbf{\tilde{d}_2} )_{s_3}
\end{align}
whose components are related via applications of the first-deformed supercharge $\tilde{\mathcal{G}}^{
\dot{\alpha} A}$. Note that the strand structure, denoted by the subscripts $s_i$, can change according to the splitting or joining action of the twist-2 operator: either a strand of winding $k$ can split into two strands of windings $k_1$, $k_2$ (with $k_1+k_2=k$), or two strands can join into a longer one. 

The difference between the CFT and the supergraviton elliptic genus for $N=2$ in \eqref{eq:genusn2} suggests there are $+20$ non-supergraviton states with $h=1$, $j=0$ which remain 1/4-BPS as one moves off the orbifold point, i.e. they are protected. Table \ref{tab:N2h1j0} shows there is a sign cancellation between the fermionic and bosonic non-supergraviton states, indicating that some non-supergravitons are lifted. 
One naively expects all supergravitons to remain protected as one moves towards the gravity regime from the free orbifold point. In the $N=2$ $(h,j)=(1,0)$ sector, it is natural to suppose the states with $\mathbf{d_2}=3$ to form the following quartets:
\begin{align} \label{eq:longmults}
\begin{split}
    &(0, \mathbf{1},\mathbf{3},\mathbf{1})_{1} + (\tfrac{1}{2}, \mathbf{2},\mathbf{3},\mathbf{1})_{2} + (1, \mathbf{1},\mathbf{3},\mathbf{1})_{1}\, ,\\[2pt]
    &(1, \mathbf{1},\mathbf{3},\mathbf{1})_{1} + (\tfrac{3}{2}, \mathbf{2},\mathbf{3},\mathbf{1})_{2} + (2 , \mathbf{1},\mathbf{3},\mathbf{1})_{1} \, ,\\[2pt]
    &(1, \mathbf{2},\mathbf{3},\mathbf{1})_{1r} + (\tfrac{3}{2}, \mathbf{1},\mathbf{3},\mathbf{1})_{2r}+ (\tfrac{3}{2}, \mathbf{1},\mathbf{3},\mathbf{1})_{2r} + (2 , \mathbf{2},\mathbf{3},\mathbf{1})_{1r}\,. \\[2pt]
\end{split}
\end{align}
One therefore naively expects the protected states to consist of the supergravitons and the $20$ fortuitous states
\begin{equation}\label{eq:fort}
   |F_U\rangle =  L_{-\frac{1}{2}} J^-_{\frac{1}{2}} |\dot{A}\dot{B}\rangle_2 \quad \text{and} \quad  |F_T\rangle = L_{-\frac{1}{2}} J^-_{\frac{1}{2}} |r\rangle_2.
\end{equation}
However, there are three different states in Table \ref{tab:N2h1j0} with the representation $(1, \mathbf{1},\mathbf{3},\mathbf{1})_{1}$ and it is not guaranteed that the graviton gas state 
\begin{equation}
    \tau^i_{\dot{A}\dot{C}}\varepsilon_{\dot{B}\dot{D}} J^{-}_{0} |\dot{A}\dot{B} \rangle^{(1)}_1  \otimes |\dot{C}\dot{D} \rangle^{(2)}_1 + \Big[(1) \leftrightarrow (2) \Big]
\end{equation}
does not appear in the long multiplets (\ref{eq:longmults}), despite the naive expectation. It is of interest to resolve this ambiguity by explicitly calculating the action of the deformed supercharge on states in Table \ref{tab:N2h1j0} to determine the precise form of the lifted and protected states. Note that the marginal deformation (\ref{eq:margdef}) we consider does not mix states in different $Z_2$ twist sectors and thus it is consistent to restrict ourselves to the $Z_2$ untwisted sector. In other words, we are free to focus specifically on the lifting of $(1, \mathbf{1},\mathbf{3},\mathbf{1})_{1}$ states.

At $N=3$, the agreement between the CFT and supergraviton indices at $(h,j)=(1,0)$ suggests that the supergravitons are protected and all non-supergravitons form long multiplets. The $N=2$ protected states in \eqref{eq:fort} do not carry over to $N=3$, consistently with their fortuitous nature. The $(\tilde{j}, \mathbf{d_1},\mathbf{d_2}, \mathbf{\tilde{d}_2} )_s$ quantum numbers of the states in Table \ref{tab:N3h1j0} show that this is indeed possible, but there is again an ambiguity of which states enter into which quartets. In particular, non-supergravitons and supergravitons may mix to form (un)lifted states. There is also an additional $(1, \mathbf{1},\mathbf{3},\mathbf{1})_{3}$ state, which may mix with the three $(1, \mathbf{1},\mathbf{3},\mathbf{1})_{1}$ states. 

\section{Multiplet recombination in perturbation theory}
\label{sec:def}

As mentioned in the previous section, we are interested in the action of the deformed supercharge (\ref{eq:deformG}) on states at the free orbifold point, in order to identify precisely those that are lifted and those that are protected. For the $N=2$ theory, the first-order deformation of the right-moving supercharges can be implemented using the $\rho$-basis formalism introduced in \eqref{eq:r2rho}-\eqref{eq:Sigmapp}. The first order deformation $\delta \mathcal{\tilde G}^{\pm A}$ of the free supercurrent ${\tilde G}^{\pm A}$ can be written as \cite{Gava:2002xb}
\begin{equation}\label{eq:deltaGrho}
    \begin{aligned}
      \delta \mathcal{\tilde G}^{\pm A}(z,\bar z) &= \oint_z dw\, G^{-A}(w)\, \Sigma_2^{+\pm}(z,\bar z) \\
      &= \mathds{1}|_{\rho=0}\otimes \left(-i\,\tau^{A\dot 1}_{\rho=1} \,e^{\frac{i}{2}(H-K)_{\rho=1}}+\tau^{A\dot 2}_{\rho=1} \,e^{\frac{i}{2}(K-H)_{\rho=1}}\right) \left(\tilde \sigma_{\rho=1} \,e^{\pm\frac{i}{2}(\tilde{H}+\tilde{K})_{\rho=1}}\right) \\
      &+ (\rho=0 \leftrightarrow \rho=1)\;,
      \end{aligned}
\end{equation}
where the symmetrisation $\rho=0 \leftrightarrow \rho=1$ is motivated as explained after~\eqref{eq:ermbosdef} and $\tau^{A \dot A}_\rho$ is the excited bosonic twist field of dimension $h=\frac{3}{4}$:
\begin{equation}
    \partial X^{A \dot A}_{\rho}(z)\,\sigma_{\rho}(0) \sim z^{-\frac{1}{2}}\,\tau^{A \dot A}_\rho(0) \,.
\end{equation}
To compute the action of the supercharge $\delta \mathcal{\tilde G}^{+ A}_{-1/2}$ on an operator at $z=0$ one should select the term of order $z^{-1}\,{\bar z}^0$ in the OPE of $\delta \mathcal{\tilde G}^{+ A}(z,\bar z)$ with the operator in question; analogously, $\delta \mathcal{\tilde G}^{- A}_{+1/2}$ is given by the term of order $z^{-1}\,{\bar z}^{-1}$ in the OPE with $\delta \mathcal{\tilde G}^{- A}(z,\bar z)$ divided by $i$, where this latter factor is needed to ensure the hermiticity relation between $\delta \mathcal{\tilde G}^{+ A}_{-1/2}$ and $\delta \mathcal{\tilde G}^{- A}_{+1/2}$.

We focus on the lifting of states with $(h,j)=(1,0)$, starting from the case $N=2$, shown in Table \ref{tab:N2h1j0}. The states with $(\tilde{j}, \mathbf{d_1},\mathbf{d_2}, \mathbf{\tilde{d}_2} ) = (1, \mathbf{1},\mathbf{1}, \mathbf{1})$ and $(1, \mathbf{1},\mathbf{3}, \mathbf{3})$, listed in the first row of that table, are ``singletons'' obtained by acting with $J^-_0$ on chiral primary operators. The $(1, \mathbf{1},\mathbf{1}, \mathbf{3})$ supergraviton state in the first row is not a singleton but there are no other states with the appropriate quantum numbers to form a quartet with this state. Thus, one expects all these three states to be protected and one can indeed check that they are annihilated by $\delta \mathcal{\tilde G}^{\pm A}_{\mp 1/2}$. Of primary interest here is the $(1, \mathbf{1},\mathbf{3}, \mathbf{1})$ sector, where there is an ambiguity related to which two of the three states listed in Table \ref{tab:N2h1j0} are lifted and which one is protected. 

We are in particular interested in how the lifting of $(1, \mathbf{1},\mathbf{3}, \mathbf{1} )$ states changes at general $N> 2$. It is convenient to introduce the following basis of $(h,j)=(1,0)$ states: states are labeled based on their right-moving R-charge $\tilde{j}$, the number of
non-trivial strands, $p$, which plays the role of the particle number in the dual gravity picture, and the winding, $w$, of the longest strand. Such a state will be denoted by $|\tilde{j}, p\rangle_w$
if it has unit norm, whilst the non-normalized state is denoted  $||\tilde{j}, p\rangle_w$. 

States where all non-trivial strands are supergravitons (i.e. chiral primaries or their left-moving global descendants) are the naive representatives of the so-called ``graviton-gas'' and will be denoted by $|{\tilde j},\mathrm{g.g.}\rangle_w$. When it it useful to highlight the dependence on $N$, we will do so by adding a subscript, for example $\|{\tilde j},p\rangle_{w,N}$; unless explicitly indicated, all states are assumed to be residing within the $S_N$ theory.
The basis of states for the lowest values of $\tilde j$ is 
\begin{equation}\label{eq:states0}
\|0,1\rangle_1^{i}=\tau^i_{\dot A \dot B}\,\sum_{r}^N \psi^{-\dot A}_{-1/2 (r)}\,\psi^{+\dot B}_{-1/2 (r)}|0\rangle~,\quad|0,1\rangle_1^{i}=\frac{1}{\sqrt{N}}\,\|0,1\rangle_1^{i} \,;
\end{equation}
\begin{equation}\label{eq:stateshalf}
\begin{aligned}
\|1/2,1\rangle_2^{i,A} & =\tau^i_{\dot A \dot B}\,\sum_{r\not = s}^N (a^{A \dot A}_{-1/2}\, \psi^{-\dot B}_{0} \,
|\Sigma^{++}_2\rangle)_{(rs)} \prod_{t\not=r,s} |0\rangle_{(t)}\,\,,\,
\\ |1/2,1\rangle_2^{i,A} & =\frac{1}{\sqrt{2 N(N-1)}}\,\|1/2,1\rangle_2^{i,A}\,; 
\end{aligned}
\end{equation}
\begin{align}
\|1,\mathrm{g.g.}\rangle_1^i&=\tau^i_{\dot A \dot B} \,\varepsilon_{\dot C \dot D}\,\sum_{r\not=s}^N\psi^{-\dot A}_{-1/2(r)}\,\psi^{+\dot B}_{-1/2 (s)} \,{\tilde \psi}^{+\dot C}_{-1/2 (r)} \,{\tilde \psi}^{+\dot D}_{-1/2 (s)}|0\rangle\,, \nonumber \\ 
|1,\mathrm{g.g.}\rangle_1^i&=\frac{1}{\sqrt{2N(N-1)}}\,\|1,\mathrm{g.g.}\rangle_1^i\,, \nonumber \\ \label{eq:states1}
\|1,1\rangle_1^i&=\tau^i_{\dot A \dot B}\,\sum_r^N \psi^{-\dot A}_{-1/2 (r)}\, \psi^{+\dot B}_{-1/2 (r)}\,{\tilde \psi}^{+\dot 1}_{-1/2 (r)} \,{\tilde \psi}^{+\dot 2}_{-1/2 (r)}|0\rangle~,\quad|1,1\rangle_1^i=\frac{1}{\sqrt{N}}\,\|1,1\rangle_1^i \,,\\ \nonumber
\|1,2\rangle_1^i &=\tau^i_{\dot A \dot B}\, \sum_{r\not=s}^N \psi^{-\dot A}_{-1/2 (r)}\, \psi^{+\dot B}_{-1/2 (r)} \,{\tilde \psi}^{+\dot 1}_{-1/2 (s)} \,{\tilde \psi}^{+\dot 2}_{-1/2 (s)}|0\rangle\;,\,\,|1,2\rangle_1^i=\frac{1}{\sqrt{N(N-1)}}\,\|1,2\rangle_1^i\,,\\ \nonumber
\|1,1\rangle_3^i
&=\tau^i_{\dot A \dot B}\,\sum_{r\not = s\not = t}^N (\psi^{-\dot A}_{-1/6} \,
\psi^{-\dot B}_{+1/6} \,|\Sigma^{++}_3\rangle)_{(rst)}\prod_{u\not=r,s,t} |0\rangle_{(u)} \,\,,\\ \nonumber \,|1,1\rangle_3^i&=\frac{1}{\sqrt{3 N(N-1)(N-2)}}\,\|1,1\rangle_3^i\,;
\end{align}
\begin{equation}\label{eq:states3half}
\begin{aligned}
\|3/2,1\rangle_2^{i,A}
&=\tau^i_{\dot A \dot B}\, \sum_{r\not = s}^N (a^{A \dot A}_{-1/2}\, \psi^{-\dot B}_{0}\,{\tilde \psi}^{+\dot 1}_{-1/2} \,{\tilde \psi}^{+\dot 2}_{-1/2}  \,|\Sigma^{++}_2\rangle)_{(rs)} \prod_{t\not=r,s} |0\rangle_{(t)}\,,\\
|3/2,1\rangle_2^{i,A} &=\frac{1}{\sqrt{2 N(N-1)}}\,\|3/2,1\rangle_2^{i,A}\,,\\
\|3/2,2\rangle_2^{i,A}&=\tau^i_{\dot A \dot B}\,\sum_{r\not = s \not=t}^N(a^{A \dot A}_{-1/2} \psi^{-\dot B}_{0} |\Sigma^{++}_2\rangle)_{(rs)}\,{\tilde \psi}^{+\dot 1}_{-1/2 (t)} \,{\tilde \psi}^{+\dot 2}_{-1/2 (t)} |0\rangle_{(t)} \prod_{u\not=r,s,t} |0\rangle_{(u)}\,,\\
|3/2,2\rangle_2^{i,A}&=\frac{1}{\sqrt{2 N(N-1)(N-2)}}\,\|3/2,2\rangle_2^{i,A}\,.
\end{aligned}
\end{equation}
Here $|0\rangle$ represents the SL$(2,C)$ invariant vacuum of the full orbifold theory, while $|0\rangle_{(r)}$ is the vacuum of the $r^{\rm th}$ copy of the seed CFT and $|\Sigma_w^{++}\rangle_{(r_1\ldots r_w)}$ is the state with twist $w$ created by the action of the CPO with $h=\bar{h}=(w-1)/2$; 
the bosonic modes $a^{A \dot A}_n=\alpha^{A\dot A}_n/\sqrt{|n|}$ satisfy the unit-normalized commutator $[a^{A\dot A}_{-n},a^{B \dot B}_{n}]=\varepsilon^{AB}\varepsilon^{\dot A \dot B}$ ($n>0$).  The symmetric matrices $\tau^i_{\dot A \dot B}$ ($i=1,2,3$) are 
\begin{equation} \label{eq:paulimat}
    \tau^1 =\frac{1}{\sqrt{2}}\,\begin{pmatrix} 1 & 0 \\ 0 & -1  \end{pmatrix}\quad,\quad \tau^2 =\frac{1}{\sqrt{2}}\,\begin{pmatrix} 1 & 0 \\ 0 & 1  \end{pmatrix}\quad,\quad \tau^3 =-\frac{1}{\sqrt{2}}\,\begin{pmatrix} 0 & 1 \\ 1 & 0  \end{pmatrix}\,.
\end{equation}
This sequence of states can be straightforwardly extended to larger values of $\tilde j$ by adding strands of winding one occupied by $\tilde J^+_{-1} ={\tilde \psi}^{+\dot 1}_{-1/2} \,{\tilde \psi}^{+\dot 2}_{-1/2}$.

We determine the action of  $\delta \mathcal{\tilde G}^{\pm A}_{\mp 1/2}$ on states for generic $N$ by applying the following series of arguments. 
\begin{itemize}
\item  Firstly, from general combinatoric arguments we deduce the action of $\delta \mathcal{\tilde G}^{\pm A}_{\mp 1/2}$ on the basis of the non-normalized states introduced in\eqref{eq:states0}-\eqref{eq:states3half}, up to $N$-independent numerical coefficients. For example, when $\delta \mathcal{\tilde G}^{+ A}_{- 1/2}$ acts on $\|1,2\rangle_1^i$, it can either join the two non-trivial strands, giving a state proportional to $\|3/2,1\rangle_2^{i,A}$, or the non-trivial strand with $\tilde j=0$ with a vacuum strand, producing 
a multiple of $\|3/2,1\rangle_2^{i,A}$; thus, one expects
\begin{equation}
 \mathcal{\tilde G}^{+ A}_{- 1/2} \, \|1,2\rangle_1^i = c_1\,\|3/2,1\rangle_2^{i,A}+ c_2\,\|3/2,2\rangle_2^{i,A}\,.
\end{equation}
The coefficients $c_1$ and $c_2$ do not depend on $N$ because the number of terms associated with the sums on the copy indices $r,s,\ldots$ has the same scaling with $N$ on both sides of the equation (the scaling is $N(N-1)$ for the $c_1$ term and $N(N-1)(N-2)$ for the $c_2$ term). Note that the naive graviton gas state, $\|1,\mathrm{g.g.}\rangle_1^i$, has the special property that it is annihilated by the deformed supercharges when they pair a non-trivial and a vacuum strand, since individual supergraviton strands are protected. A qualitatively different structure appears for the state $\|1,1\rangle_3^i$. Here, the joining action of the supercharges is trivial, because there are no states of winding four for $h=1$, $j=0$. So $\delta \mathcal{\tilde G}^{+ A}_{- 1/2}$ can only split the winding three strand into a winding two and a winding one strand. When the singly wound strand is non-trivial, one generates a state proportional to $\|3/2,2\rangle_2^{i,A}$ and the scaling with $N$ works as before. But one could also produce a trivial singly wound strand, resulting in a state proportional to $\|3/2,1\rangle_2^{i,A}$; the number of copies contained in this state scales like $N(N-1)$, while 
the scaling of $\mathcal{\tilde G}^{+ A}_{- 1/2}\,\|1,1\rangle_3^i$ is $N(N-1)(N-2)$. This implies that 
\begin{equation}\label{eq:c3c4}
\mathcal{\tilde G}^{+ A}_{- 1/2}\,\|1,1\rangle_3^i = c_3\,(N-2)\,\|3/2,1\rangle_2^{i,A}+c_4\,\|3/2,2\rangle_2^{i,A}\,,
\end{equation}
with some $N$-independent coefficients $c_3$ and $c_4$. Note that the production of a vacuum strand is the only general mechanism that can generate $N$-dependent coefficients when the deformed supercharges act on the non-normalized states defined above. 
\item Some of the $N$-independent coefficients $c_i$ can be computed for $N=2$ using the $\rho$-basis form of the supercharges, \eqref{eq:deltaGrho}. The details of this computation are given in appendix~\ref{sec:appA}.
\item Coefficients like $c_3$ and $c_4$ in \eqref{eq:c3c4}, which are associated with states that are trivial for $N=2$, do not follow from the calculation above. A worldsheet derivation of such coefficients requires the orbifold techniques of \cite{Guo:2019ady,Guo:2020gxm,Guo:2022ifr,Hughes:2023apl,Hughes:2023fot}. However, in our case, general consistency requirements are enough to determine, and actually over-constrain, the missing coefficients. We impose two types of constraints: (i) The deformed supercharges must satisfy the anti-commutators
\begin{equation}
\{\mathcal{\tilde G}^{+ A}_{- 1/2},\mathcal{\tilde G}^{- B}_{1/2} \}=\varepsilon^{AB}\,{\tilde h}\,,
\end{equation}
where ${\tilde h}$ is the right-moving anomalous dimension, that has to be the same for all the states in the same quartet. (ii) Since different quartets have, generically, different anomalous dimensions, states belonging to different multiplets must be orthogonal to each other.
\end{itemize}
The final outcome of this procedure is
\begin{equation}\label{eq:Gpm0}
\delta\mathcal{\tilde G}^{+A}_{-1/2} \,\|0,1\rangle_1^i = \frac{1}{\sqrt{2}}\,\|1/2,1\rangle_2^{i,A}\quad,\quad \delta\mathcal{\tilde G}^{-A}_{1/2} \,\|0,1\rangle_1^i = 0\,,
\end{equation}
\begin{equation}
\begin{aligned}\label{eq:Gpmhalf}
\delta\mathcal{\tilde G}^{+A}_{-1/2} \,\|1/2,1\rangle_2^{i,B} &=\varepsilon^{AB}\,\frac{1}{\sqrt{2}}\,\left(\|1,\mathrm{g.g.}\rangle_1^i +(N-1)\, \|1,1\rangle_1^i +\|1,2\rangle_1^{i} -\|1,1\rangle_3^{i} \right)\\
&\equiv \varepsilon^{AB}\,\frac{1}{\sqrt{2}}\,\|1,\mathrm{top}\rangle^i \,,\\
\delta\mathcal{\tilde G}^{-A}_{1/2} \,\|1/2,1\rangle_2^{i,B} &=\sqrt{2} \,(N-1)\,\varepsilon^{AB}\,\|0,1\rangle^i_1\,,
\end{aligned}
\end{equation}
\begin{equation}\label{eq:Gp1}
\begin{aligned}
\delta\mathcal{\tilde G}^{+A}_{-1/2} \,\|1,\mathrm{g.g.}\rangle_1^i&=\frac{1}{\sqrt{2}}\,\|3/2,1\rangle_2^{i,A}\,,\\
\delta\mathcal{\tilde G}^{+A}_{-1/2} \,\|1,1\rangle_1^{i}&= -\frac{1}{2\sqrt{2}}\|3/2,1\rangle_2^{i,A}\,,\\
\delta\mathcal{\tilde G}^{+A}_{-1/2} \,\|1,2\rangle_1^{i}&= -\frac{1}{2\sqrt{2}}\|3/2,1\rangle_2^{i,A} + 
\frac{1}{\sqrt{2}}\, \|3/2,2\rangle_2^{i,A}\,,\\
\delta\mathcal{\tilde G}^{+A}_{-1/2} \,\|1,1\rangle_3^{i}&= -\frac{N-2}{2\sqrt{2}}\,\|3/2,1\rangle_2^{i,A} + \frac{1}{\sqrt{2}}\,\|3/2,2\rangle_2^{i,A}  \,,
\end{aligned}
\end{equation}
\begin{equation}\label{eq:Gm1}
    \begin{aligned}
\delta\mathcal{\tilde G}^{-A}_{1/2} \,\|1,\mathrm{g.g.}\rangle_1^{i} &=-\frac{1}{\sqrt{2}}\,\|1/2,1\rangle_2^{i,A}\,,\\
\delta\mathcal{\tilde G}^{-A}_{1/2} \,\|1,1\rangle_1^{i} &=-\frac{1}{2\sqrt{2}}\,\|1/2,1\rangle_2^{i,A}\,,\\
\delta\mathcal{\tilde G}^{-A}_{1/2} \,\|1,2\rangle_1^{i} &=-\frac{1}{2\sqrt{2}}\,\|1/2,1\rangle_2^{i,A} \,,\\
\delta\mathcal{\tilde G}^{-A}_{1/2} \,\|1,1\rangle_3^{i} &=\frac{3}{2\sqrt{2}}\,(N-2)\,\|1/2,1\rangle_2^{i,A} \,.
\end{aligned}
\end{equation}

Knowing the explicit action of the supercharges on the states allow us to derive the structure of the multiplets. The four states with $\tilde j=1$ can be combined into three non-protected and one protected state. The non-protected states are: the state $\|1,\mathrm{top}\rangle^i$, defined in \eqref{eq:Gpmhalf}
\begin{align}\label{eq:top}
    \|1,&\mathrm{top}\rangle^i = \left(\|1,\mathrm{g.g.}\rangle_1^i +(N-1)\, \|1,1\rangle_1^i +\|1,2\rangle_1^{i} -\|1,1\rangle_3^{i} \right)  \\ \nonumber
    &  = \sqrt{N(N-1)}\,\left(\sqrt{2} \, |1,\mathrm{g.g.}\rangle_1^i +\sqrt{N-1}\, |1,1\rangle_1^i +|1,2\rangle_1^{i} -\sqrt{3(N-2)}|1,1\rangle_3^{i}\right)
\end{align}
which forms the top component ($\mathcal{\tilde G}^{+A}_{-1/2} \,\|1,\mathrm{top}\rangle^i=0$) of the quartet starting with $\|0,1\rangle_1^i$; the two states 
\begin{gather} \label{eq:bottom1}
  \| 1,\mathrm{bottom}_1\rangle^{i}=\|1,\mathrm{g.g.}\rangle_1^i -(N-1)\, \|1,1\rangle_1^i + \|1,2\rangle_1^{i} -\frac{N-4}{3(N-2)} \,\|1,1\rangle_3^{i} = \\ \nonumber
  \sqrt{N(N-1)} \left[  \sqrt{2} |1,\mathrm{g.g.}\rangle_1^i - \sqrt{(N-1)} \, |1,1\rangle_1^i + |1,2\rangle_1^{i} -\frac{N-4}{\sqrt{3(N-2)}}\,|1,1\rangle_3^{i}\ \right] \,,\\ \label{eq:bottom2}
\|1,\mathrm{bottom}_2\rangle^{i}=\|1,\mathrm{g.g.}\rangle_1^i -(N-1)\, \|1,1\rangle_1^i +(3-2N)\, \|1,2\rangle_1^{i} - \|1,1\rangle_3^{i} = \\ \nonumber
                                   \sqrt{N(N-1)} \left[  \sqrt{2} |1,\mathrm{g.g.}\rangle_1^i - \sqrt{(N-1)} \, |1,1\rangle_1^i + (3-2N)\, |1,2\rangle_1^{i} -\sqrt{3(N-2)}\,|1,1\rangle_3^{i}\ \right] \,,
\end{gather}
that are the bottom components ($\mathcal{\tilde G}^{-A}_{1/2} \,\|1,\mathrm{bottom}_{1,2}\rangle^i = 0 $) of two orthogonal quartets.
Finally, the unique protected state ($\mathcal{\tilde G}^{+A}_{-1/2}\,\|1,\mathrm{singlet}\rangle^i=\mathcal{\tilde G}^{-A}_{1/2} \,\|1,\mathrm{singlet}\rangle^i = 0 $) at this level is
\begin{align} \label{eq:singlet}
&\|1,\mathrm{singlet}\rangle^i = \|1,\mathrm{g.g.}\rangle_1^i +\frac{N-1}{N-2}\, \|1,1\rangle_1^i -\frac{1}{N-2}\, \|1,2\rangle_1^{i} +\frac{1}{N-2} \,\|1,1\rangle_3^{i}\\ \nonumber
&=\sqrt{N(N-1)}\,\left[\sqrt{2}\,|1,\mathrm{g.g.}\rangle_1^i+\frac{\sqrt{N-1}}{N-2}\,|1,1\rangle_1^i -\frac{1}{N-2}\, |1,2\rangle_1^{i} +\frac{\sqrt{3}}{\sqrt{N-2}} \,|1,1\rangle_3^{i}
\right]\,.
\end{align}
The expressions in terms of the normalized basis make the large $N$ interpretation of each state evident: in Eqs.~\eqref{eq:top} and~\eqref{eq:bottom1} the dominant terms have only a single strand excited and so should be identified as stringy states on the bulk side, instead Eqs.~\eqref{eq:bottom2} and~\eqref{eq:singlet} should be interpreted as multi-particle states. Note that the protected state~\eqref{eq:singlet} coincides with the naive supergraviton state, $\|1,\mathrm{g.g.}\rangle_1^i$, in the large $N$ limit and so it can be interpreted holographically as a bound state of two $1/2$-BPS objects. However, its finite $N$ expression contains sub-leading $1/N$ corrections necessary to make the state closed with respect to $Q^A \equiv \mathcal{\tilde G}^{+A}_{-1/2}$ and $Q^{A\,\dagger}\equiv \mathcal{\tilde G}^{-A}_{1/2}$. A similar pattern is found for $1/4$-BPS operators in $\mathcal{N}=4$ SYM. For example, the $1/4$-BPS operator in the $[2,0,2]$ rep found in \cite{Ryzhov:2001bp,DHoker:2001jzy} contains a naive double-trace term and a $1/N$ single-trace correction that is needed to make the state $Q^{A\,\dagger}$ closed at one loop. 

The limit $N=2$ has to be treated separately because some of the structures appearing in the general results of~\eqref{eq:Gp1} do not exist. In this case the protected state that takes the place of~\eqref{eq:singlet} is
\begin{equation}
  \label{eq:singlN2}
   \|1,\mathrm{singlet}\rangle^i_{N=2} = \|1,1\rangle_{1,2}^{i} - \|1,2\rangle_{1,2}^{i} = \sqrt{2} \left[|1,1\rangle_{1,2}^{i} - |1,2\rangle_{1,2}^{i}\right]\,
\end{equation}
and there are only two non-protected states
\begin{equation}
  \label{eq:liftN2}
  \begin{aligned}
         \|1,\mathrm{top}\rangle^i_{N=2} &=\|1,\mathrm{g.g.}\rangle_{1,2}^i + \|1,1\rangle_{1,2}^{i} + \|1,2\rangle_{1,2}^{i} \\ &= 2 |1,\mathrm{g.g.}\rangle_{1,2}^i + \sqrt{2}|1,1\rangle_{1,2}^{i} + \sqrt{2}|1,2\rangle_{1,2}^{i}\;, \\
         \|1,\mathrm{bottom}\rangle^i_{N=2} &=\|1,\mathrm{g.g.}\rangle_{1,2}^i -\|1,1\rangle_{1,2}^{i} - \|1,2\rangle_{1,2}^{i}\,\\ &= 2 |1,\mathrm{g.g.}\rangle_{1,2}^i -\sqrt{2}|1,1\rangle_{1,2}^{i} - \sqrt{2}|1,2\rangle_{1,2}^{i}\;.
  \end{aligned}
\end{equation}

A non-trivial check on the form of the state in \eqref{eq:singlet} can be performed using the strong coupling data on protected three-point functions. Consider the correlator between the normalized state
\begin{equation}
|1,\mathrm{singlet}\rangle^i = \frac{N-2}{\sqrt{2 N(N-1)}(N-1)}\,\|1,\mathrm{singlet}\rangle^i 
\end{equation}
and two single-particle $1/2$-BPS operators. This class of correlators is protected \cite{Baggio:2012rr} and thus its value computed in the free theory should agree with the strong coupling result, which can be extracted from the holographic four-point function of single-particle $1/2$-BPS states derived in \cite{Giusto:2018ovt,Ceplak:2021wzCz}. The $1/2$-BPS operators of dimension $(1/2,1/2)$, $O^{\alpha,\dot\alpha}_f$ ($\alpha,\dot\alpha=\pm$, $f=1,\ldots 21$), form a vector of the $SO(21)$ flavor symmetry of the supergravity theory compactified on K3. Their supergravity four-point correlator in the R-symmetry configuration that is relevant for us is \cite{Giusto:2018ovt}:
\begin{equation}\label{eq:s1111}
\begin{aligned}
    \langle O^{++}_{f_1}(0) O^{--}_{f_2}(\infty) O^{+-}_{f_3}(1) O^{-+}_{f_4}(z,\bar z)\rangle &= \frac{\delta_{f_1 f_2}\delta_{f_3 f_4}}{(1-z)(1-\bar z)} \left[1-\frac{1}{N}+ \frac{\bar z}{N}\, {\bar D}_{1 1 2 2} \right] \\
    &+ \frac{\delta_{f_1 f_3} \delta_{f_2 f_4}}{N}\,{\bar z}\,{\bar D}_{1212}+\frac{\delta_{f_1 f_4} \delta_{f_2 f_3}}{N\,z}\,{\bar D}_{2112}\,,
    \end{aligned}
\end{equation}
with ${\bar D}_{\Delta_1 \Delta_2 \Delta_3 \Delta_4}$ the usual $\bar D$ functions of holographic correlators. The state $|1,\mathrm{singlet}\rangle^i$ is exchanged at leading order in the $z\to 0$ OPE channel of the above four-point function anti-symmetrized\footnote{We define $O^{++}_{[f_1} O^{-+}_{f4]} = \frac{1}{2} (O^{++}_{f_1} O^{-+}_{f4}-O^{++}_{f_4} O^{-+}_{f1})$.} with respect to the $(f_1,f_4)$ flavor indices:
\begin{equation}\label{eq:sugra}
\lim_{z\to 0} \langle O^{++}_{[f_1}(0) O^{--}_{f_2}(\infty) O^{+-}_{f_3}(1) O^{-+}_{f_4]}(z,\bar z)\rangle = (\delta_{f_1 f_2}\delta_{f_3 f_4}-\delta_{f_1 f_3}\delta_{f_2 f_4})\,\frac{1}{2}\left(1-\frac{1}{N}\right)\,.
\end{equation}
The $SO(21)$ flavor symmetry is generically broken at the free orbifold point but it must be realized in the protected sector, although not explicitly. The most direct way to compare the free-theory calculation with the supergravity result \eqref{eq:sugra} is to pick any two flavors, for example
\begin{equation}\label{eq:Opp}
\begin{aligned}
O^{++}_{f_1} = \frac{\tau^1_{\dot A\dot B}}{\sqrt{N}}\sum_{r=1}^N \psi^{+\dot A}_{(r)}\, {\tilde \psi}^{+\dot B}_{(r)}\quad&,\quad O^{--}_{f_2} = -\frac{\tau^1_{\dot A\dot B}}{\sqrt{N}}\sum_{r=1}^N \psi^{-\dot A}_{(r)}\, {\tilde \psi}^{-\dot B}_{(r)}\,,\\
O^{+-}_{f_3}= \frac{\tau^3_{\dot A\dot B}}{\sqrt{N}}\sum_{r=1}^N \psi^{+\dot A}_{(r)} \,{\tilde \psi}^{-\dot B}_{(r)}\quad &,\quad O^{-+}_{f_4}= \frac{\tau^3_{\dot A\dot B}}{\sqrt{N}}\sum_{r=1}^N \psi^{-\dot A}_{(r)} \,{\tilde \psi}^{+\dot B}_{(r)} \,,
\end{aligned}
\end{equation}
where the $\tau$ matrices are defined in~\eqref{eq:paulimat}. One finds
\begin{equation}\label{eq:OPEOO}
 \lim_{z\to 0}\, O^{++}_{[f_1}(z,\bar z) O^{-+}_{f_4](0)} = \frac{\sqrt{N-1}}{2\sqrt{N}}\,|1,\mathrm{g.g.}\rangle^2_1 +\frac{1}{\sqrt{2 N}}\,|1,1\rangle^2_1+\frac{\sqrt{N-1}}{2\sqrt{N}}\,|1,\mathrm{(1,1,1,3)}\rangle^2_1\,,
\end{equation}
where
\begin{equation}
   |1,\mathrm{(1,1,1,3)}\rangle^i_1 =\frac{1}{\sqrt{2 N (N-1)}} \tau^i_{\dot C \dot D} \,\varepsilon_{\dot A \dot B}\,\sum_{r\not=s}^N\psi^{-\dot A}_{-1/2(r)}\,\psi^{+\dot B}_{-1/2 (s)} \,{\tilde \psi}^{+\dot C}_{-1/2 (r)} \,{\tilde \psi}^{+\dot D}_{-1/2 (s)}\,,
\end{equation}
is the $1/4$-BPS supergraviton operator in the $(1, \mathbf{1},\mathbf{1}, \mathbf{3})$ sector. Then, the protected three-point correlators computed in the free theory are 
\begin{equation}\label{eq:free3pt}
    \langle O^{++}_{[f_1} O^{-+}_{f_4]} \,|1,\mathrm{singlet}\rangle_1^2 = \frac{1}{2}\,\sqrt{1-\frac{1}{N}}\quad,\quad \langle O^{++}_{[f_1} O^{-+}_{f_4]} \,|1,\mathrm{(1,1,1,3)}\rangle_1^2 = \frac{1}{2}\,\sqrt{1-\frac{1}{N}}\,.
\end{equation}
Note that the supersymmetric operator $|1,\mathrm{singlet}\rangle_1^i$ has exactly the same three-point coupling as the naive supergraviton state $|1,\mathrm{g.g.}\rangle^i_1$. The supergravity result \eqref{eq:sugra} is due to the exchange of the two supersymmetric states, $|1,\mathrm{singlet}\rangle_1^i$ and $|1,(1, \mathbf{1},\mathbf{1}, \mathbf{3})\rangle^i_1$, and it is indeed equal to the sum of the squares of the two three-point functions \eqref{eq:free3pt}. 

It is interesting to compare the OPE limits of the strong coupling and the free theory four-point functions. At the orbifold point, all the three operators that appear on the r.h.s. of \eqref{eq:OPEOO} are supersymmetric and are exchanged in the corresponding four-point correlator. The $z\to 0$ OPE limit of the free four-point function should thus be given by the sum of the squares of the three three-point couplings:
\begin{equation}\label{eq:OPEfree}
    \frac{N-1}{4 N}+\frac{1}{2N}+\frac{N-1}{4 N}=\frac{1}{2}\,.
\end{equation}
This differs from the strong coupling result in \eqref{eq:sugra} because of the $1/N$ term. We can check this OPE prediction by computing the free correlator of the operators in \eqref{eq:Opp}: 
\begin{equation}\label{eq:s1111free}
    \langle O^{++}_{[f_1}(0) O^{--}_{f_2}(\infty) O^{+-}_{f_3}(1) O^{-+}_{f_4]}(z,\bar z)\rangle_\mathrm{free} =\frac{1}{2} \left[\frac{1}{|1-z|^2} + \frac{1}{N}\,\frac{\bar{z}}{z(1-\bar{z})}\right]\;.
\end{equation}
Note that the anti-symmetrized correlator is now singular in the $z\to 0$ limit, signaling the lack of flavor symmetry in the free theory, but the term of order $z^0\, {\bar z}^0$ reproduces exactly \eqref{eq:OPEfree}.

\section{Revisiting the notion of fortuity and discussion}
\label{sec:comm}

It is interesting to compare the spectrum of theories with different values of $N$. In order to do so, one needs to define a projection operator $\Pi$ mapping states of the $N+1$ theory into those of the $N$ theory. Once $\Pi$ is defined for any $N$, it is of course possible to connect theories with any $N''>N'$ with subsequent applications of $\Pi$. The most natural definition of the projection~\cite{Chang:2025rqy} is given in terms of the non-normalised states in the symmetric orbifold CFT introduced in the previous section: when acting on states in the $N+1$ theory that include a non-zero number of strands of winding one in the NSNS vacuum, the projection gives back the state with one fewer such strands, which is naturally part of the $N$ theory; instead the projection is trivial when acting on states that do not have vacuum strands of winding one. This operation is equivalent to simply reducing the range of the sums in \eqref{eq:states0}-\eqref{eq:states3half} from $N+1$ to $N$. Following the notation of~\cite{Chang:2025rqy}, we will refer to this projection operator with the symbol $\Pi^{\rm pre}$, to stress that it is a preliminary definition that will be refined. For instance, acting on the states defined in~\eqref{eq:states0} we have
\begin{equation}
  \label{eq:Pipre}
  \Pi^{\rm pre} \,\|0,1\rangle_{1,N+1}^{i} = \|0,1\rangle_{1,N}^{i}\;,\quad \Pi^{\rm pre}\,\|1/2,1\rangle^{i,A}_{2,N+1} = \|1/2,1\rangle^{i,A}_{2,N}\,.
\end{equation}
Note that by identifying the $S_N$ Hilbert space with the subset of the $S_{N+1}$ states containing at least one vacuum strand, one could think of $\Pi^{\rm pre}$ as a map on the $S_{N+1}$ Hilbert space that satisfies the standard projector condition $(\Pi^{\rm pre})^2=\Pi^{\rm pre}$. This justifies the name commonly used in the literature.

The projection operator introduced above does not in general commute with the global generators of the superalgebra~\cite{Chang:2025rqy}. For instance, in the orbifold Sym$_N(T^4)$, there are extra $U(1)$ generators and one needs to dress $\Pi^{\rm pre}$ with appropriate insertions to ensure that they commute with the full projection operator. Since here we focus on the case Sym$_N(K3)$, which does not have extra $U(1)$ generators, we will not need to take this issue into account. Instead we have to be careful with the failure of $\Pi^{\rm pre}$ to commute with the supercharges~\cite{Chang:2025rqy}. For instance we have
\begin{equation}
  \label{eq:failcomQ}
  [\Pi^{\rm pre},\mathcal{\tilde G}^{+A}_{-1/2} ] \, \|1/2,1\rangle^{i,B}_{2,N+1} = \sqrt{2}\,\varepsilon^{AB}\,\|1,1\rangle_{1,N}^i\,,
\end{equation}
which follows from the $N$-dependent term in~\eqref{eq:Gpmhalf}. Because of this, the sequence of BPS-states~\eqref{eq:singlet} is not compatible with the naive projection operator:
\begin{equation}
  \label{eq:Pipresingl}
  \Pi^{\rm pre} \|1,\mathrm{singlet}\rangle^i_{N+1} \not= \|1,\mathrm{singlet}\rangle^i_N\,.
\end{equation}

A similar failure of the commutation relation between the supercharge and the projection operator is present in the SYK two-flavour model of~\cite{Heydeman:2022lse}, as pointed out in~\cite{Chang:2024lxt}. The proposal of~\cite{Chang:2024lxt,Chang:2025rqy} is to still use $\Pi^{\rm pre}$ to define monotone sequences. According to this approach, we have (for $N>3$)
  \begin{equation}
    \label{eq:Pipreseq}
    \begin{aligned}
      (\Pi^{\rm pre})^{-1} \|1,\mathrm{singlet}\rangle^i_{N-1} & = \|1,\mathrm{g.g.}\rangle_{1,N}^i +\frac{N-2}{N-3}\, \|1,1\rangle_{1,N}^i \\ & -  \frac{1}{N-3}\, \|1,2\rangle_{1,N}^{i} +\frac{1}{N-3} \,\|1,1\rangle_{3,N}^{i} + \|\alpha\rangle^i_{N}\,,
      \end{aligned}
  \end{equation}
where $\|\alpha\rangle_{N}^i$ is an arbitrary state on which $\Pi^{\rm pre}$ acts trivially. The issue is that~\eqref{eq:Pipreseq} is not in the same cohomology class of~\eqref{eq:singlet} for any allowed $\|\alpha \rangle_N^i$ and so one would draw the conclusion that the singlet state~\eqref{eq:singlet} is not part of a monotone sequence. This is in tension with the factorisation of the 4-point correlator at strong and weak coupling discussed at the end of Sect.~\ref{sec:def} which shows that there exists a protected state with $(h,j)=(1,0)$ in the $(1, \mathbf{1},\mathbf{3}, \mathbf{1})$ irrep for any value of $N$.

In order to restore the pattern expected for a sequence of monotone states, we propose that the definition of the projection map should be modified so as it commutes with the supercharges. We will work only at first order in conformal perturbation theory and focus on the sector $(h,j)=(1,0)$ as done throughout this paper. However this is already instructive as it highlights a general pattern. As in~\eqref{eq:failcomQ}, the failure of the commutation between $\Pi^{\rm pre}$ and $\mathcal{\tilde G}^{+A}_{-1/2}$ always originates from the $N$-dependent terms in the action of the supercharge on the non-normalised states. Such terms arise from the splitting action of the supercharge and, in particular, when it acts on a multi-wound strand yielding a product of a trivial singly wound strand and an excited strand, see the discussion around \eqref{eq:c3c4}.\footnote{This is very similar to what happens in the SYK two-flavour model, where, contrary to what happens for the original SYK model, the action of the supercharge can produce copies of the Hilbert space which do not contain any excitations~\cite{Chang:2024lxt}.} This suggest that the action of the projection map on strands that can be split by the supercharges in the fashion mentioned above should be revisited. We will do so by assuming that in these cases the result is an arbitrary linear combinations of the state produced by $\Pi^{\rm pre}$ and all other states with the same quantum numbers.

Then there is no room for any change in~\eqref{eq:Pipre} since there are no other states with $\tilde{j}=0$ or $\tilde{j}=1/2$. The first modification suggested by the discussion above appears at $\tilde{j}=1$ and can be parameterised as follows 
\begin{equation}
  \label{eq:Piw3}
  \begin{aligned}
    & \Pi \,\|1,\ldots\rangle_{1,N+1}^i = \Pi^{\rm pre} \,\|1,\ldots\rangle_{1,N+1}^i\;,\\ &\Pi \,\|1,1\rangle_{3,N+1}^i =x_0\,\|1,\mathrm{g.g.}\rangle^i_{1,N} + x_1\,\|1,1\rangle^i_{1,N}+x_2\,\|1,2\rangle^i_{1,N}+x_3\,\|1,1\rangle^i_{3,N}\,.
    \end{aligned}
\end{equation}
The action of $\Pi^{\rm pre}$ corresponds to setting all $x_i$ to zero, except $x_3$ which is one. Instead we fix these parameters by requiring that $[\Pi,\mathcal{\tilde G}^{+A}_{-1/2} ]$ vanishes when acting on states of increasing values of $\tilde{j}$. The first constraint follows from the case discussed in~\eqref{eq:failcomQ} which singles out $x_0=x_2=0$ and $x_1=x_3=1$, so we have
\begin{equation}\label{eq:Piw4}
\Pi \,\|1,1\rangle_{3,N+1}^i = \,\|1,1\rangle^i_{1,N} + \,\|1,1\rangle^i_{3,N}\,.
\end{equation}
This automatically ensures that the commutator vanishes on the states with $\tilde{j}=1$ by taking $\Pi\,\|3/2,\ldots\rangle^{i,A}_{2,N+1} = \Pi^{\rm pre}\,\|3/2,\ldots\rangle^{i,A}_{2,N+1}$. At least in the sector of interest here the pattern continues at higher values of $\tilde{j}$: it is sufficient to modify the action of the projection map on states containing a strand of winding three, adding a term with an excited strand of winding one and two extra vacuum strand. With appropriate $N$-independent coefficients, that depend on the explicit action of the supercharges, it is possible to ensure that $[\Pi,\mathcal{\tilde G}^{+A}_{-1/2} ]$ vanishes when acting on all states with $(h,j)=(1,0)$. It would of course be interesting to study other sectors to make sure that a similar upgrade of $\Pi^{\rm pre}$ is always possible. Note that if one thinks of $\Pi$ as a map on the $S_{N+1}$ Hilbert space, in the sense explained above, it fails to satisfy the projector condition: $\Pi^2\not=\Pi$. We keep referring to it as a projection map to highlight its relation with usual projector used in the fortuity literature. 

The refined projection map should link monotone cohomology classes in a natural way. Let us check that this is the case for the family identified in~\eqref{eq:singlet}. By using the results above, it is straightforward to check the following identity for $N>2$
\begin{align}
  \Pi\,\|1,\mathrm{singlet}\rangle^i_{N+1} & =  \|1,\mathrm{g.g.}\rangle_{1,N}^i +\frac{N+1}{N-1}\, \|1,1\rangle_{1,N}^i +\frac{1}{N-1}\left( -\|1,2\rangle_{1,N}^{i} + \,\|1,1\rangle_{3,N}^{i}\right)\nonumber \\
    \label{eq:chk1} & = \frac{N(N-2)}{(N-1)^2} \|1,\mathrm{singlet}\rangle^i_N + \frac{1}{(N-1)^2} \|1,\mathrm{top}\rangle^i_N\,.
\end{align}
Since the second term on the r.h.s. is the top component of a quartet, it is an exact term for both $\mathcal{\tilde G}^{+A}_{-1/2}$ supercharges and so $\Pi\,\|1,\mathrm{singlet}\rangle^i_{N+1}$ is in the same cohomology class as $\|1,\mathrm{singlet}\rangle^i_N$. 

Notice that this result can also be extended to $N=2$ by taking the formal\footnote{It is important to keep into account that the state $ \|1,\mathrm{singlet}\rangle^i_N$ in \eqref{eq:singlet} contains terms that diverge for $N\to 2$; this divergence is canceled by $N-2$ factor in \eqref{eq:chk1}.} $N\to 2$ limit of the first term in the second line of~\eqref{eq:chk1} and by discarding at the same time any strand of winding larger than $2$. It is then clear that the projection of the $N=3$ singlet state to $N=2$ is in the same cohomology class as~\eqref{eq:singlN2}, which completes the monotone to the lowest values of $N$ even if the explicit form of~\eqref{eq:singlN2} is counterintuitive since it does not include any graviton gas component. The analysis of~\cite{Hughes:2026naj} suggests that this is a generic feature of various protected states in the $N=2$ case.

As anticipated in Sect.~\ref{sec:symmo}, in the $N=2$ theory there are fortuitous states with $(h,j)=(1,0)$. Let us consider the $Z_2$ untwisted state in~\eqref{eq:fort}: in the basis~\eqref{eq:r2rho} the corresponding vertex operator reads as 
\begin{align}
  \label{eq:RSstate2}
   O_{F_U} & = 
   \Big[\psi^{+\dot{A}}_{\rho=0} \tilde\psi^{+\dot{B}}_{\rho=0}\otimes \left(\sigma_{\rho=1} \,e^{-\frac{i}{2}(H+K)_{\rho=1}} \tilde \sigma_{\rho=1} \,e^{\frac{i}{2}(\tilde{H}+\tilde{K})_{\rho=1}}\right) \\ & + \psi^{-\dot{A}}_{\rho=0} \tilde\psi^{+\dot{B}}_{\rho=0}\otimes \left(\sigma_{\rho=1} \,e^{\frac{i}{2}(H+K)_{\rho=1}} \tilde \sigma_{\rho=1} \,e^{\frac{i}{2}(\tilde{H}+\tilde{K})_{\rho=1}}\right) \Big] + (\rho=0 \leftrightarrow \rho=1) \,. \nonumber
\end{align}
This state is in general part of a long multiplet because the action of $\mathcal{\tilde G}^{+A}_{-1/2}$ is non-trivial. However, for $N=2$, the contribution proportional to $\|3/2,2\rangle_2^{i,A}$ is absent and so the state is accidentally protected (since the terms proportional to $\|3/2,1\rangle_2^{i,A}$ cancel). We can apply the same technique discussed in Appendix \ref{sec:appA} to calculate the OPE between the fortuitous state~\eqref{eq:RSstate2} and the monotone state $\Sigma^{+-}$ (actually the latter is even more particular, as it is a single-particle supergraviton state in the multiplet of~\eqref{eq:Sigmapp}). The leading term in this OPE is another supergraviton state
\begin{equation}
    \label{MMF}
    O_{F_U} (z,\bar{z}) \Sigma^{+-}(0,0) \sim \frac{1}{z \bar{z}} \sum_r \psi^{+\dot{A}}_{(r)} \tilde\psi^{+\dot{B}}_{(r)} + \ldots\;.
\end{equation}
This shows that, at least in the $N=2$ case, there is no superselection rule separating dynamically monotone and fortuitous states. If this were the case in general, this would imply that, when there exist both fortuitous and monotone states, it would not be justified to study the two ensembles separately. For example, when investigating the presence of chaos \cite{Shenker:2013pqa} with the criterion suggested in \cite{Lin:2022rzw,Lin:2022zxd,Chen:2024oqv} one should presumably have to consider a full matrix comprising both monotone and fortuitous states. For instance, it has recently been pointed out~\cite{Chen:2026vml} that ad-hoc restrictions on the set of states can alter the results of the criteria used to diagnose chaos. 

It would be very interesting to study how this mixing between monotone and fortuitous states behaves as the central charge is increased. It is possible that the three-point functions with a single fortuitous state and two monotone ones, such as the one mentioned above, get exponentially suppressed in $N$. However, it is unclear that this would be sufficient to decouple the two classes of states in the gravity regime, since the number of fortuitous states is expected to be exponentially large, so even exponentially small couplings can lead to a sizeable mixing. Another indication that it is hard to separate monotone and fortuitous states is the existence of regular solutions with an arbitrarly long AdS$_2$ throat~\cite{Bena:2016ypk} -- a feature that one naively associates with fortuitous states -- but an explicit CFT dual in the monotone class~\cite{Bena:2017xbt}. This issue certainly deserves further study, starting from the precise form of the dual CFT state,  which may receive corrections with respect to the naive form along the lines of what we saw in this paper for the state~\eqref{eq:singlet}. 

Another direction which is worthwhile to investigate is how to precisely formulate questions about the large $N$ limit in a way that is relevant for black hole physics. In this paper we focused on a sector with fixed quantum numbers $(h,j)=(1,0)$ and studied how fortuitous and monotone states changed as $N$ increases. For applications to black holes, it is more natural to increase $h$ and $N$ at the same time keeping fixed the curvature scale of the backreacted geometry. To be concrete, for the superstrata that are coherent superposition of gravitons, this means increasing $N$ by keeping fixed the parameters determining the coherent states. The precise form of the dual states at the orbifold point is a key ingredient of the precision holography program started long ago, see~\cite{Kanitscheider:2006zf,Kanitscheider:2007wq} for a discussion in the D1D5/AdS$_3$ case which was then extended to the superstrata geometries in~\cite{Giusto:2015dfa,Giusto:2019qig,Rawash:2021pik}. The analysis of this paper suggests that, in the large $N$ scaling mentioned above, there may be new contributions that are non-linear in the coherent parameters, something that could be tested at strong coupling extending the precision holography analysis.

Of course the most interesting black holes are not supersymmetric and so one may suspect that the distinction between monotone and fortuitous states may be relevant also beyond the BPS case where it has been developed so far. For instance, one could ask whether the top elements of quartets at different values of $N$ form sequences related by the map $\Pi$. This is indeed the case for the state~\eqref{eq:top}
\begin{equation}
    \label{eq:topproj}
    \Pi \|1,\mathrm{top}\rangle^i_{N+1} = \|1,\mathrm{top}\rangle^i_N\;,
\end{equation}
as it is easy to check by using~\eqref{eq:Piw3} and~\eqref{eq:Piw4}. As mentioned, the holographic interpretation of this family of states is in terms of a single-particle stringy mode and so it is natural that it exists for all values of $N$. We expect that in general there will be other non-BPS states that do not form similar sequences. For example, the distinction between primary and secondary invariants for operators constructed from hermitean matrices has many features in common with the fortuity mechanism without relying on supersymmetry~\cite{deMelloKoch:2026utx}.

Finally we saw that, in order to define monotone states with the expected properties, it is important to restore the commutation between the projection operator and the supercharges $Q$. In the D1D5 case, this commutation is broken already by the first corrections to the free theory, but in other holographic theories this may happen only at higher orders in conformal perturbation theory. This may be the case of ${\cal N}=4$ SYM and then considerations similar to those discussed in this paper will apply there as well.

\subsection*{Acknowledgements}
We would like to thank Marcel Hughes and Masaki Shigemori for several discussions and for collaboration in a related project. 
RR would like to thank Haoyu Zhang for discussions, Matthias Gaberdiel for an illuminating converation on the $T^4/Z_2$ symmetric orbifold and Robert de Mello Koch for very useful feedback on a preliminary version of this work. JI and RR would like to thank James Chryssanthacopoulos and David Vegh for patient explanations on fortuity in the SYK models. SG is grateful to Camillo Imbimbo for useful comments on the supersymmetry algebra. It is a pleasure to thank Tarek Anous, Samir Mathur and Ida Zadeh for several enlighting discussions over the years on closely related topics.
RR is supported by the UK EPSRC grant ``CFT and Gravity: Heavy States and Black Holes" EP/W019663/1 and by the Science and Technology Facilities Council (STFC) Consolidated Grant ST/X00063X/1 ``Amplitudes, Strings \& Duality''. No new data were generated or analysed during this study.

\appendix
    \section{Calculating Coefficients for $N=2$ Case} \label{sec:appA}
The numerical coefficients in Equation (\ref{eq:Gpm0}) can be derived as follows. 
Using the basis of states (\ref{eq:states0}), we can write the $(h,j)=(1,0)$ state with $\tilde{j}=0$ in the $\rho$-basis (\ref{eq:r2rho}) as
\begin{equation} 
    \|0,1\rangle_1^{i}=\lim_{z\rightarrow0} \tau^i_{\dot A \dot B}\,\Big( \psi^{-\dot A}_{(\rho=0)}(z)\,\psi^{+\dot B}_{(\rho=0) } (z)+\psi^{-\dot A}_{(\rho=1)}(z)\,\psi^{+\dot B}_{(\rho=1)}(z) \Big)  |0\rangle\;.
\end{equation}
Consider the polarization
\begin{equation}
    \tau^{i=+}_{\dot A \dot B} = \frac{1}{\sqrt{2}} (\tau^1_{\dot A \dot B} + \tau^2_{\dot A \dot B} )\;.
\end{equation}
The bosonization relations (\ref{eq:ermbosdef}) can be used to write the state in terms of the auxiliary bosons $H$ and $K$
\begin{equation} \label{eq:coeff1}
     \|0,1\rangle_1^{i}=\lim_{z\rightarrow0}\Big( ie^{i(K-H)_{\rho=0}(z)}  + ie^{i(K-H)_{\rho=1}(z)} \Big)  |0\rangle\;.
\end{equation}
In order to compute the action of $\delta \mathcal{\tilde G}^{+ A}_{-1/2}$ (\ref{eq:deltaGrho}) on the state, one needs to select the $(w-z)^{-1}(\bar{w}-\bar{z})^0$ term in the OPE between $\delta \mathcal{\tilde G}^{+ A}(w,\bar{w})$ and the operator in (\ref{eq:coeff1}) located at $(z,\bar{z})$. We have
\begin{equation} \label{eq:ope1}
\begin{aligned} 
& \delta \mathcal{\tilde G}^{+ A}(w, \bar{w})  \,  \Big( ie^{i(K-H)_{\rho=0}(z)}  + ie^{i(K-H)_{\rho=1}(z)} \Big)   \\[2mm]
&\sim \frac{1}{w-z}
\Bigg[
\mathds{1}_{\rho=0}\otimes
\tau^{A\dot 1}_{\rho=1}\,
e^{-\frac{i}{2}(H-K)_{\rho=1}}
\Big(
\tilde{\sigma}_{\rho=1}\,
e^{\frac{i}{2}(\tilde H+\tilde K)_{\rho=1}}
\Big)+(\rho=0 \leftrightarrow \rho=1)\Bigg] 
+\cdots \;,
\end{aligned}
\end{equation}
where the dots denote regular terms. The operator in the square brackets corresponds to the state produced when acting $\delta \mathcal{\tilde G}^{+ 2}_{-1/2}$ on the state $\|0,1\rangle_1^{i=+}$. The final state contains two terms, each which have norm-squared equal to one, giving the overall state a norm-squared of $2$. In order to identify this final state with the state $\|1/2,1\rangle_2^{i,2}$ in Equation (\ref{eq:stateshalf}), one needs to introduce a numerical factor to ensure the norms match when swapping from the $\rho$ to $r$-basis. The state $\|1/2,1\rangle_2^{i,A}$ has norm-squared equal to $4$, and one needs to introduce a numerical factor of $1/\sqrt{2}$:
\begin{equation} 
    \delta\mathcal{\tilde G}^{+A}_{-1/2} \,\|0,1\rangle_1^i = \frac{1}{\sqrt{2}}\,\|1/2,1\rangle_2^{i,A}.
\end{equation}
If we instead consider the action of $\delta \mathcal{\tilde G}^{- A}_{1/2}$ on the state, one needs to select the $(w-z)^{-1}(\bar{w}-\bar{z})^{-1}$ term in the OPE between $\delta \mathcal{\tilde G}^{- A}(w,\bar{w})$ and the operator in question:
\begin{equation} \label{eq:ope2}
\begin{aligned} 
 \delta \mathcal{\tilde G}^{- A}(w, \bar{w})  \, \Big( ie^{i(K-H)_{\rho=0}(z)}  + ie^{i(K-H)_{\rho=1}(z)} \Big)  \,.
\end{aligned}
\end{equation}
It is not possible to obtain a $(\bar{w}-\bar{z})^{-1}$ term in this OPE and we can therefore conclude that $\delta \mathcal{\tilde G}^{- A}_{1/2}$ annihilates the state,
\begin{equation}
    \mathcal{\tilde G}^{-A}_{1/2} \,\|0,1\rangle_1^i = 0\;.
\end{equation}

The numerical coefficients in Equation (\ref{eq:Gpmhalf}) can be derived as follows. We first write the state 
\begin{equation} 
\begin{aligned} 
    \|1/2,1\rangle_2^{i,A}=\lim_{z\rightarrow0}\tau^i_{\dot A \dot B}\, \bigg(2\,(a^{A \dot A}_{-1/2}\, \psi^{-\dot B}_{0} \, \Sigma^{++}_2)_{(12)} (z) \bigg) |0\rangle\;.
\end{aligned}
\end{equation}
This state has norm-squared of four. In the $\rho$-basis, considering the case where $\tau^i = \tau^+$ and $A=2$, this state is written as
\begin{align} \nonumber 
    \|1/2,1\rangle_2^{i,2}&= \lim_{z\rightarrow0}\sqrt{2} \Big[ \mathds{1}_{\rho=0} \otimes     \left(\tau^{2\dot{1}}_{\rho=1} \,e^{\frac{i}{2}(K-H)_{\rho=1} (z)}\right) \left(\tilde \sigma_{\rho=1} \,e^{\frac{i}{2}(\tilde{H}+\tilde{K})_{\rho=1}(z)}\right) + \ldots \Big] |0\rangle\;,
\end{align}
where the dots stand for the terms symmetrising the expression in the exchange $(\rho=0 \leftrightarrow \rho=1)$. The $\sqrt{2}$ factor is included in order to reproduce the same norm-squared. Consider the action of $\delta \mathcal{\tilde G}^{+ A}_{-1/2}$ (\ref{eq:deltaGrho}) on this state. The OPE between $\delta \mathcal{\tilde G}^{+ 1}$ and the operator above with $A=2$ is given by
\begin{equation}\label{eq:Gt1n2}
\begin{aligned} 
& \delta \mathcal{\tilde G}^{+ 1} (w, \bar{w}) \, \times \sqrt{2} \Big[ \mathds{1}_{\rho=0} \otimes    \left(\tau^{2\dot{1}}_{\rho=1} \,e^{\frac{i}{2}(K-H)_{\rho=1}(z)}\right) \left(\tilde \sigma_{\rho=1} \,e^{\frac{i}{2}(\tilde{H}+\tilde{K})_{\rho=1}(z)}\right) + \ldots \Big]  \\ &
\sim-\frac{\sqrt{2}}{w-z}\Bigg[
\mathds{1}_{\rho=0}\otimes
\Big(
e^{i(K-H)_{\rho=1}}
\Big)
\Big(
e^{i(\tilde H+\tilde K)_{\rho=1}}
\Big)+
(\rho=0\leftrightarrow\rho=1)
\Bigg]  + \ldots\;,
\end{aligned}
\end{equation}
where the dots in the second line denote regular terms. To compare with \eqref{eq:Gpmhalf} it is useful to rewrite the $\tilde j=1$ terms in the $\rho$ basis and in the bosonized language:
\begin{equation} 
    \|1,\mathrm{g.g.}\rangle_1^{i}= \lim_{z\rightarrow0}\Big( ie^{i(K-H)_{\rho=0}(z)}  - ie^{i(K-H)_{\rho=1}(z)} \Big)\Big( ie^{i(\tilde H+\tilde K)_{\rho=0}(\bar{z})} -ie^{i(\tilde H+\tilde K)_{\rho=1}(\bar{z})}\Big)|0\rangle\,,
\end{equation}
\begin{equation} 
    \|1,1\rangle_1^{i}+ \|1,2\rangle_1^{i}=\,\lim_{z\rightarrow0}\Big( ie^{i(K-H)_{\rho=0}(z)}  + ie^{i(K-H)_{\rho=1}(z)} \Big)\Big( ie^{i(\tilde H+\tilde K)_{\rho=0}(\bar{z})} +ie^{i(\tilde H+\tilde K)_{\rho=1}(\bar{z})}\Big)|0\rangle\,.
\end{equation} 
Thus
\begin{equation} 
   \|1,\mathrm{g.g.}\rangle_1^{i}+  \|1,1\rangle_1^{i}+ \|1,2\rangle_1^{i}=-2\lim_{z\rightarrow0}\Big[\Big(
e^{i(K-H)_{\rho=0}}
\Big)
\Big(
e^{i(\tilde H+\tilde K)_{\rho=0}}
\Big)+
(\rho=0\leftrightarrow\rho=1) \Big]|0\rangle\,.
\end{equation}
Therefore, comparing with \eqref{eq:Gt1n2}, we have for $N=2$
\begin{equation}
    \delta\mathcal{\tilde G}^{+A}_{-1/2} \,\|1/2,1\rangle_2^{i,B} =\varepsilon^{AB}\,\frac{1}{\sqrt{2}}\,\left(\|1,\mathrm{g.g.}\rangle_1^i + \|1,1\rangle_1^i +\|1,2\rangle_1^{i}  \right)\;.
\end{equation}
The action of $\delta \mathcal{\tilde G}^{-1}_{1/2}$ on the state is given by the singular term in the OPE
\begin{equation} 
\begin{aligned} 
&
\delta \mathcal{\tilde G}^{- 1} (w, \bar{w}) \times
    \sqrt{2} \Big[ \mathds{1}_{\rho=0} \otimes     \left(\tau^{2\dot{1}}_{\rho=1} \,e^{\frac{i}{2}(K-H)_{\rho=1} (z)}\right) \left(\tilde \sigma_{\rho=1} \,e^{\frac{i}{2}(\tilde{H}+\tilde{K})_{\rho=1} (z)}\right) + (\rho=0 \leftrightarrow \rho=1)\ \Big] \\
    &
\sim \frac{-\sqrt{2}}{(w-z)(\bar{w}-\bar{z})}\Bigg[
\mathds{1}_{\rho=0}\otimes
\Big(
e^{i(K-H)_{\rho=1}}
\Big) 
+
(\rho=0\leftrightarrow\rho=1)
\Bigg]  + \ldots .
\end{aligned}
\end{equation}
Using the recipe given after \eqref{eq:deltaGrho}, one can conclude that
\begin{equation}
    \delta\mathcal{\tilde G}^{-A}_{1/2} \,\|1/2,1\rangle_2^{i,B} =\varepsilon^{AB}\,\sqrt{2}\,\|0,1\rangle_1  \,.
\end{equation}

\providecommand{\href}[2]{#2}\begingroup\raggedright\endgroup


\begin{thebibliography}{10}

\bibitem{Maldacena:1997re}
J.~M. Maldacena, ``{The large N limit of superconformal field theories and
  supergravity},'' {\em Adv. Theor. Math. Phys.} {\bf 2} (1998) 231--252,
\href{http://arXiv.org/abs/hep-th/9711200}{{\tt hep-th/9711200}}.

\bibitem{Shenker:2013pqa}
S.~H. Shenker and D.~Stanford, ``{Black holes and the butterfly effect},'' {\em
  JHEP} {\bf 03} (2014) 067, \href{http://arXiv.org/abs/1306.0622}{{\tt
  1306.0622}}.

\bibitem{Schlenker:2022dyo}
J.-M. Schlenker and E.~Witten, ``{No ensemble averaging below the black hole
  threshold},'' {\em JHEP} {\bf 07} (2022) 143,
  \href{http://arXiv.org/abs/2202.01372}{{\tt 2202.01372}}.

\bibitem{Chen:2024oqv}
Y.~Chen, H.~W. Lin, and S.~H. Shenker, ``{BPS chaos},'' {\em SciPost Phys.}
  {\bf 18} (2025), no.~2, 072, \href{http://arXiv.org/abs/2407.19387}{{\tt
  2407.19387}}.

\bibitem{Lin:2004nb}
H.~Lin, O.~Lunin, and J.~M. Maldacena, ``{Bubbling AdS space and 1/2 BPS
  geometries},'' {\em JHEP} {\bf 10} (2004) 025,
\href{http://arXiv.org/abs/hep-th/0409174}{{\tt hep-th/0409174}}.

\bibitem{Lunin:2001fv}
O.~Lunin and S.~D. Mathur, ``{Metric of the multiply wound rotating string},''
  {\em Nucl. Phys.} {\bf B610} (2001) 49--76,
\href{http://arXiv.org/abs/hep-th/0105136}{{\tt hep-th/0105136}}.

\bibitem{Kanitscheider:2007wq}
I.~Kanitscheider, K.~Skenderis, and M.~Taylor, ``{Fuzzballs with internal
  excitations},'' {\em JHEP} {\bf 06} (2007) 056,
  \href{http://arXiv.org/abs/0704.0690}{{\tt 0704.0690}}.

\bibitem{Gutowski:2004ez}
J.~B. Gutowski and H.~S. Reall, ``{Supersymmetric AdS(5) black holes},'' {\em
  JHEP} {\bf 02} (2004) 006,
\href{http://arXiv.org/abs/hep-th/0401042}{{\tt hep-th/0401042}}.

\bibitem{Gutowski:2004yv}
J.~B. Gutowski and H.~S. Reall, ``{General supersymmetric AdS(5) black
  holes},'' {\em JHEP} {\bf 04} (2004) 048,
\href{http://arXiv.org/abs/hep-th/0401129}{{\tt hep-th/0401129}}.

\bibitem{Strominger:1996sh}
A.~Strominger and C.~Vafa, ``{Microscopic Origin of the Bekenstein-Hawking
  Entropy},'' {\em Phys. Lett.} {\bf B379} (1996) 99--104,
\href{http://arXiv.org/abs/hep-th/9601029}{{\tt hep-th/9601029}}.

\bibitem{Bena:2015bea}
I.~Bena, S.~Giusto, R.~Russo, M.~Shigemori, and N.~P. Warner, ``{Habemus
  Superstratum! A constructive proof of the existence of superstrata},'' {\em
  JHEP} {\bf 05} (2015) 110,
\href{http://arXiv.org/abs/1503.01463}{{\tt 1503.01463}}.

\bibitem{Bena:2017xbt}
I.~Bena, S.~Giusto, E.~J. Martinec, R.~Russo, M.~Shigemori, D.~Turton, and
  N.~P. Warner, ``{Asymptotically-flat supergravity solutions deep inside the
  black-hole regime},'' {\em JHEP} {\bf 02} (2018) 014,
\href{http://arXiv.org/abs/1711.10474}{{\tt 1711.10474}}.

\bibitem{Shigemori:2020yuo}
M.~Shigemori, ``{Superstrata},'' {\em Gen. Rel. Grav.} {\bf 52} (2020), no.~5,
  51, \href{http://arXiv.org/abs/2002.01592}{{\tt 2002.01592}}.

\bibitem{Bena:2016ypk}
I.~Bena, S.~Giusto, E.~J. Martinec, R.~Russo, M.~Shigemori, D.~Turton, and
  N.~P. Warner, ``{Smooth horizonless geometries deep inside the black-hole
  regime},'' {\em Phys. Rev. Lett.} {\bf 117} (2016), no.~20, 201601,
\href{http://arXiv.org/abs/1607.03908}{{\tt 1607.03908}}.

\bibitem{Chang:2024zqi}
C.-M. Chang and Y.-H. Lin, ``{Holographic covering and the fortuity of black
  holes},'' \href{http://arXiv.org/abs/2402.10129}{{\tt 2402.10129}}.

\bibitem{Chang:2013fba}
C.-M. Chang and X.~Yin, ``{1/16 BPS states in $\mathcal N=$ 4 super-Yang-Mills
  theory},'' {\em Phys. Rev. D} {\bf 88} (2013), no.~10, 106005,
  \href{http://arXiv.org/abs/1305.6314}{{\tt 1305.6314}}.

\bibitem{Chang:2022mjp}
C.-M. Chang and Y.-H. Lin, ``{Words to describe a black hole},'' {\em JHEP}
  {\bf 02} (2023) 109, \href{http://arXiv.org/abs/2209.06728}{{\tt
  2209.06728}}.

\bibitem{Choi:2022caq}
S.~Choi, S.~Kim, E.~Lee, and J.~Park, ``{The shape of non-graviton operators
  for SU(2)},'' {\em JHEP} {\bf 09} (2024) 029,
  \href{http://arXiv.org/abs/2209.12696}{{\tt 2209.12696}}.

\bibitem{Chang:2024lxt}
C.-M. Chang, Y.~Chen, B.~S. Sia, and Z.~Yang, ``{Fortuity in SYK models},''
  {\em JHEP} {\bf 08} (2025) 003, \href{http://arXiv.org/abs/2412.06902}{{\tt
  2412.06902}}.

\bibitem{Belin:2025hsg}
A.~Belin, P.~Singh, R.~Vadala, and A.~Zaffaroni, ``{Fortuity in ABJM},''
  \href{http://arXiv.org/abs/2512.04146}{{\tt 2512.04146}}.
  
\bibitem{Behan:2025hbx}
C.~Behan and L.~P.~de Gioia, ``Two roads to fortuity in ABJM theory,''
{\em JHEP} {\bf 06} (2026) 217, \href{http://arXiv.org/abs/2512.23603 }{{\tt 2512.23603}}. 

\bibitem{deMelloKoch:2025ngs}
R.~de~Mello~Koch and A.~Jevicki, ``{Structure of loop space at finite N},''
  {\em JHEP} {\bf 06} (2025) 011, \href{http://arXiv.org/abs/2503.20097}{{\tt
  2503.20097}}.

\bibitem{deMelloKoch:2025cec}
R.~de~Mello~Koch, A.~Ghosh, and H.~J.~R. Van~Zyl, ``{Bosonic fortuity in vector
  models},'' {\em JHEP} {\bf 06} (2025) 246,
  \href{http://arXiv.org/abs/2504.14181}{{\tt 2504.14181}}.

\bibitem{Chen:2025sum}
Y.~Chen, ``{Fortuity with a single matrix},''
  \href{http://arXiv.org/abs/2511.00790}{{\tt 2511.00790}}.

\bibitem{Chang:2025rqy}
C.-M. Chang, Y.-H. Lin, and H.~Zhang, ``{Fortuity in the D1-D5 system},''
  \href{http://arXiv.org/abs/2501.05448}{{\tt 2501.05448}}.

\bibitem{Hughes:2025tdy}
M.~R.~R. Hughes and M.~Shigemori, ``{Fortuity and supergravity},'' {\em JHEP}
  {\bf 03} (2026) 130, \href{http://arXiv.org/abs/2505.14888}{{\tt
  2505.14888}}.

\bibitem{Chang:2025wgo}
C.-M. Chang and H.~Zhang, ``{Fortuity and R-charge concentration in the D1-D5
  CFT},'' \href{http://arXiv.org/abs/2511.23294}{{\tt 2511.23294}}.

\bibitem{Zhang:2026jnf}
H.~Zhang, ``{Signatures of Quantum Chaos in the D1D5 System},''
  \href{http://arXiv.org/abs/2605.18725}{{\tt 2605.18725}}.

\bibitem{Baggio:2012rr}
M.~Baggio, J.~de~Boer, and K.~Papadodimas, ``{A non-renormalization theorem for
  chiral primary 3-point functions},'' {\em JHEP} {\bf 07} (2012) 137,
\href{http://arXiv.org/abs/1203.1036}{{\tt 1203.1036}}.

\bibitem{Gava:2002xb}
E.~Gava and K.~Narain, ``{Proving the PP wave / CFT(2) duality},'' {\em JHEP}
  {\bf 0212} (2002) 023,
\href{http://arXiv.org/abs/hep-th/0208081}{{\tt hep-th/0208081}}.

\bibitem{Carson:2014yxa}
Z.~Carson, S.~Hampton, S.~D. Mathur, and D.~Turton, ``{Effect of the twist
  operator in the D1D5 CFT},'' {\em JHEP} {\bf 1408} (2014) 064,
\href{http://arXiv.org/abs/1405.0259}{{\tt 1405.0259}}.

\bibitem{Carson:2014ena}
Z.~Carson, S.~Hampton, S.~D. Mathur, and D.~Turton, ``{Effect of the
  deformation operator in the D1D5 CFT},'' {\em JHEP} {\bf 01} (2015) 071,
\href{http://arXiv.org/abs/1410.4543}{{\tt 1410.4543}}.

\bibitem{Carson:2015ohj}
Z.~Carson, S.~Hampton, and S.~D. Mathur, ``{Second order effect of twist
  deformations in the D1D5 CFT},'' {\em JHEP} {\bf 04} (2016) 115,
  \href{http://arXiv.org/abs/1511.04046}{{\tt 1511.04046}}.

\bibitem{Carson:2016uwf}
Z.~Carson, S.~Hampton, and S.~D. Mathur, ``{Full action of two deformation
  operators in the D1D5 CFT},'' {\em JHEP} {\bf 11} (2017) 096,
  \href{http://arXiv.org/abs/1612.03886}{{\tt 1612.03886}}.

\bibitem{Guo:2019pzk}
B.~Guo and S.~D. Mathur, ``{Lifting of states in 2-dimensional $N = 4$
  supersymmetric CFTs},'' {\em JHEP} {\bf 10} (2019) 155,
  \href{http://arXiv.org/abs/1905.11923}{{\tt 1905.11923}}.

\bibitem{Guo:2019ady}
B.~Guo and S.~D. Mathur, ``{Lifting of level-1 states in the D1D5 CFT},'' {\em
  JHEP} {\bf 03} (2020) 028, \href{http://arXiv.org/abs/1912.05567}{{\tt
  1912.05567}}.

\bibitem{Guo:2020gxm}
B.~Guo and S.~D. Mathur, ``{Lifting at higher levels in the D1D5 CFT},'' {\em
  JHEP} {\bf 11} (2020) 145, \href{http://arXiv.org/abs/2008.01274}{{\tt
  2008.01274}}.

\bibitem{Guo:2022ifr}
B.~Guo, M.~R.~R. Hughes, S.~D. Mathur, and M.~Mehta, ``{Universal lifting in
  the D1-D5 CFT},'' {\em JHEP} {\bf 10} (2022) 148,
  \href{http://arXiv.org/abs/2208.07409}{{\tt 2208.07409}}.

\bibitem{Fiset:2022erp}
M.-A. Fiset, M.~R. Gaberdiel, K.~Naderi, and V.~Sriprachyakul, ``{Perturbing
  the symmetric orbifold from the worldsheet},'' {\em JHEP} {\bf 07} (2023)
  093, \href{http://arXiv.org/abs/2212.12342}{{\tt 2212.12342}}.

\bibitem{Hughes:2023apl}
M.~R.~R. Hughes, S.~D. Mathur, and M.~Mehta, ``{Lifting of two-mode states in
  the D1-D5 CFT},'' {\em JHEP} {\bf 01} (2024) 183,
  \href{http://arXiv.org/abs/2309.03321}{{\tt 2309.03321}}.

\bibitem{Hughes:2023fot}
M.~R.~R. Hughes, S.~D. Mathur, and M.~Mehta, ``{Lifting of superconformal
  descendants in the D1-D5 CFT},'' {\em JHEP} {\bf 04} (2024) 129,
  \href{http://arXiv.org/abs/2311.00052}{{\tt 2311.00052}}.

\bibitem{Gaberdiel:2023lco}
M.~R. Gaberdiel, R.~Gopakumar, and B.~Nairz, ``{Beyond the tensionless limit:
  integrability in the symmetric orbifold},'' {\em JHEP} {\bf 06} (2024) 030,
  \href{http://arXiv.org/abs/2312.13288}{{\tt 2312.13288}}.

\bibitem{Gaberdiel:2024nge}
M.~R. Gaberdiel, F.~Lichtner, and B.~Nairz, ``{Anomalous dimensions in the
  symmetric orbifold},'' {\em JHEP} {\bf 05} (2025) 084,
  \href{http://arXiv.org/abs/2411.17612}{{\tt 2411.17612}}.

\bibitem{Gaberdiel:2025smz}
M.~R. Gaberdiel and I.~L. Meur, ``{The triplet perturbation of the symmetric
  orbifold},'' {\em JHEP} {\bf 03} (2026) 231,
  \href{http://arXiv.org/abs/2509.03132}{{\tt 2509.03132}}.

\bibitem{Gaberdiel:2026jor}
M.~R. Gaberdiel, B.~Nairz, and C.~Peng, ``{Non-planar corrections in the
  symmetric orbifold},'' \href{http://arXiv.org/abs/2605.06465}{{\tt
  2605.06465}}.

\bibitem{Burrington:2015mfa}
B.~A. Burrington, A.~W. Peet, and I.~G. Zadeh, ``{Bosonization, cocycles, and
  the D1-D5 CFT on the covering surface},'' {\em Phys. Rev. D} {\bf 93} (2016),
  no.~2, 026004, \href{http://arXiv.org/abs/1509.00022}{{\tt 1509.00022}}.

\bibitem{Giusto:2015dfa}
S.~Giusto, E.~Moscato, and R.~Russo, ``{AdS$_{3}$ holography for 1/4 and 1/8
  BPS geometries},'' {\em JHEP} {\bf 11} (2015) 004,
\href{http://arXiv.org/abs/1507.00945}{{\tt 1507.00945}}.

\bibitem{Giusto:2019qig}
S.~Giusto, S.~Rawash, and D.~Turton, ``{Ads$_{3}$ holography at dimension
  two},'' {\em JHEP} {\bf 07} (2019) 171,
  \href{http://arXiv.org/abs/1904.12880}{{\tt 1904.12880}}.

\bibitem{Rawash:2021pik}
S.~Rawash and D.~Turton, ``{Supercharged AdS$_{3}$ Holography},'' {\em JHEP}
  {\bf 07} (2021) 178, \href{http://arXiv.org/abs/2105.13046}{{\tt
  2105.13046}}.

\bibitem{Baggio:2015jxa}
M.~Baggio, M.~R. Gaberdiel, and C.~Peng, ``{Higher spins in the symmetric
  orbifold of K3},'' {\em Phys. Rev. D} {\bf 92} (2015) 026007,
  \href{http://arXiv.org/abs/1504.00926}{{\tt 1504.00926}}.

\bibitem{Dijkgraaf:1996xw}
R.~Dijkgraaf, G.~W. Moore, E.~P. Verlinde, and H.~L. Verlinde, ``{Elliptic
  genera of symmetric products and second quantized strings},'' {\em Commun.
  Math. Phys.} {\bf 185} (1997) 197--209,
\href{http://arXiv.org/abs/hep-th/9608096}{{\tt hep-th/9608096}}.

\bibitem{deBoer:1998us}
J.~de~Boer, ``{Large N Elliptic Genus and AdS/CFT Correspondence},'' {\em JHEP}
  {\bf 05} (1999) 017,
\href{http://arXiv.org/abs/hep-th/9812240}{{\tt hep-th/9812240}}.

\bibitem{Hughes:2025oxu}
M.~R.~R. Hughes and M.~Shigemori, ``{New supersymmetry index for the D1-D5
  conformal field theories},'' {\em Phys. Rev. D} {\bf 113} (2026), no.~4,
  046022, \href{http://arXiv.org/abs/2509.19425}{{\tt 2509.19425}}.

\bibitem{Hughes:2026qqn}
M.~R.~R. Hughes and M.~Shigemori, ``{The Resolved Elliptic Genus and the D1-D5
  CFT},'' \href{http://arXiv.org/abs/2603.18138}{{\tt 2603.18138}}.

\bibitem{Ryzhov:2001bp}
A.~V. Ryzhov, ``{Quarter BPS operators in N=4 SYM},'' {\em JHEP} {\bf 11}
  (2001) 046, \href{http://arXiv.org/abs/hep-th/0109064}{{\tt hep-th/0109064}}.

\bibitem{DHoker:2001jzy}
E.~D'Hoker and A.~V. Ryzhov, ``{Three point functions of quarter BPS operators
  in N=4 SYM},'' {\em JHEP} {\bf 02} (2002) 047,
  \href{http://arXiv.org/abs/hep-th/0109065}{{\tt hep-th/0109065}}.

\bibitem{Giusto:2018ovt}
S.~Giusto, R.~Russo, and C.~Wen, ``{Holographic correlators in AdS$_{3}$},''
  {\em JHEP} {\bf 03} (2019) 096, \href{http://arXiv.org/abs/1812.06479}{{\tt
  1812.06479}}.

\bibitem{Ceplak:2021wzCz}
N.~Ceplak, S.~Giusto, M.~R.~R. Hughes, and R.~Russo, ``{Holographic correlators
  with multi-particle states},'' {\em JHEP} {\bf 09} (2021) 204,
  \href{http://arXiv.org/abs/2105.04670}{{\tt 2105.04670}}.

\bibitem{Heydeman:2022lse}
M.~Heydeman, G.~J. Turiaci, and W.~Zhao, ``{Phases of $ \mathcal{N} $ = 2
  Sachdev-Ye-Kitaev models},'' {\em JHEP} {\bf 01} (2023) 098,
  \href{http://arXiv.org/abs/2206.14900}{{\tt 2206.14900}}.

\bibitem{Hughes:2026naj}
M.~R.~R. Hughes, K.~Jin, D.~Matsumoto, L.~Miyahara, and M.~Shigemori,
  ``{Towering Gravitons in AdS$_3$/CFT$_2$},''
  \href{http://arXiv.org/abs/2604.20663}{{\tt 2604.20663}}.

\bibitem{Lin:2022rzw}
H.~W. Lin, J.~Maldacena, L.~Rozenberg, and J.~Shan, ``{Holography for people
  with no time},'' {\em SciPost Phys.} {\bf 14} (2023), no.~6, 150,
  \href{http://arXiv.org/abs/2207.00407}{{\tt 2207.00407}}.

\bibitem{Lin:2022zxd}
H.~W. Lin, J.~Maldacena, L.~Rozenberg, and J.~Shan, ``{Looking at
  supersymmetric black holes for a very long time},'' {\em SciPost Phys.} {\bf
  14} (2023), no.~5, 128, \href{http://arXiv.org/abs/2207.00408}{{\tt
  2207.00408}}.

\bibitem{Chen:2026vml}
Y.~Chen, S.~Colin-Ellerin, O.~Mamroud, and K.~Papadodimas, ``{Chaos of Berry
  curvature for BPS microstates},'' \href{http://arXiv.org/abs/2604.23287}{{\tt
  2604.23287}}.

\bibitem{Kanitscheider:2006zf}
I.~Kanitscheider, K.~Skenderis, and M.~Taylor, ``{Holographic anatomy of
  fuzzballs},'' {\em JHEP} {\bf 04} (2007) 023,
  \href{http://arXiv.org/abs/hep-th/0611171}{{\tt hep-th/0611171}}.

\bibitem{deMelloKoch:2026utx}
R.~de~Mello~Koch and J.~P. Rodrigues, ``{Secondary invariants and
  non-perturbative states},'' \href{http://arXiv.org/abs/2604.15600}{{\tt
  2604.15600}}.

\end{thebibliography}
\end{document}